\documentclass[aps,prd,onecolumn,preprintnumbers,nofootinbib, amsmath,amssymb,superscriptaddress]{revtex4}

\usepackage{graphicx}
\usepackage{dco
lumn}
\usepackage{bm,graphicx,hyperref,subfigure}
\usepackage{float,amsfonts,graphicx}
\usepackage{color}
\usepackage{tikz}
\usepackage{adjustbox}
\usetikzlibrary{decorations.pathreplacing}
\usetikzlibrary{arrows.meta}
\usepackage{placeins}
\usepackage{graphicx}
\usepackage{dcolumn}
\usepackage{bm}
\usepackage{amsmath}
\usepackage{cancel}
\usepackage{comment}
\usepackage{xcolor}

\newcommand{\bs}{\mathbf}
\newcommand{\beq}{\begin{eqnarray}}
\newcommand{\eeq}{\end{eqnarray}}

\newcommand{\real}{{\sf I}\kern-.12em{\sf R}}
\newcommand{\comp}{{\sf I}\kern-.50em{\sf C}}
\newcommand{\unity}{{\sf I}\kern-.54em{\sf 1}}

\usepackage{hyperref}

\newcommand{\tr}{\mbox{Tr}}

\newcommand{\Orsay}{IJCLab, P\^ole Th\'eorie (Bat.~210), CNRS/IN2P3 et Universit\'e,\\ Paris-Saclay, 91405 Orsay, France}
\newcommand{\Romatre}{Dipartimento di Matematica e Fisica, Universit\`a  Roma Tre and INFN, Sezione di Roma Tre,\\ Via della Vasca Navale 84, I-00146 Rome, Italy}
\newcommand{\RomatreINFN}{Istituto Nazionale di Fisica Nucleare, Sezione di Roma Tre,\\ Via della Vasca Navale 84, I-00146 Rome, Italy}
\newcommand{\Romadue}{Dipartimento di Fisica and INFN, Universit\`a di Roma ``Tor Vergata",\\ Via della Ricerca Scientifica 1, I-00133 Roma, Italy}

\begin{document}

\title{Lattice QCD determination of the radiative decay rates $h_{c}\to \eta_{c}\, \gamma$ and $h_{b}\to \eta_{b}\, \gamma$} 

\author{D.\,Be\v{c}irevi\'c}\affiliation{\Orsay}
\author{R.\,Di\,Palma}\affiliation{\Romatre}
\author{R.\,Frezzotti}\affiliation{\Romadue} 
\author{G.\,Gagliardi}\affiliation{\Romatre}
\author{V.\,Lubicz}\affiliation{\Romatre} 
\author{F.\,Sanfilippo}\affiliation{\RomatreINFN}
\author{N.\,Tantalo}\affiliation{\Romadue}

\date{\today}

\begin{abstract}

We present the results of our lattice QCD computation of the hadronic matrix elements relevant to the $h_{c}\to \eta_{c}\gamma$ and $h_{b}\to \eta_{b}\gamma$ decays by using the gauge configurations produced by the Extended Twisted Mass Collaboration with $N_{f}=2+1+1$ dynamical Wilson-Clover twisted mass fermions at five different lattice spacings with physical dynamical $u$ , $d$, $s$ and $c$ quark masses (except for the the coarsest lattice for which the lightest sea quark corresponds to a pion with $m_{\pi}\simeq 175~\mathrm{MeV}$). While the hadronic matrix element for $h_{c}\to \eta_{c}\gamma$ is obtained directly, the one relevant to $h_{b}\to\eta_{b}\gamma$ is reached by working with heavy quark masses $m^{(n)}_{H} = \lambda^{n-1} m_{c}$, with $\lambda \sim 1.24$ and $n=1,2, \ldots ,6$, and then extrapolated to $m_{b}$ by several judicious ansätze. In the continuum limit we obtain $\Gamma( h_{c}\to \eta_{c} \gamma ) = 0.604(24)~\mathrm{MeV}$, which is by a factor of $2.3$ more accurate than the previous lattice estimates, and in good agreement with the experimental measurement. In the $b$-quark case we obtain $\Gamma( h_{b}\to \eta_{b} \gamma) =46.0(4.8)~\mathrm{keV}$.

\end{abstract}

\maketitle

\section{Introduction}
In recent years, the radiative decays of heavy quarkonia have attracted considerable interest, particularly with the experimental detection of the $\eta_b$ meson by BaBar~\cite{BaBar:2008dae}. The study of the so called electric dipole (E1) and magnetic dipole (M1) transitions, such as $h_{c(b)} \rightarrow \eta_{c(b)} \gamma$, $J/\Psi \to \eta_{c}\gamma$ and $\Upsilon(nS) \to \eta_{b}\gamma$, can give further insight in the internal structure and dynamics of heavy quarkonium states conveniently described by a suitable effective field theory~\cite{Bodwin:1994jh,Brambilla:1999xf,Fleming:2000ib,Brambilla:2004jw}.

The experimental challenge in measuring decay processes with limited phase space, such as $\Upsilon(1S) \rightarrow \eta_b \gamma$, has led experimenters to focus on decays of the radially excited states, such as $\Upsilon(2S) \rightarrow \eta_b \gamma$ and $\Upsilon(3S) \rightarrow \eta_b \gamma$~\cite{Belle:2018aht, CLEO:2009nxu,BaBar:2009xir, BaBar:2008dae}. These studies helped extracting $m_{\eta_b}$ and provided us with estimates of the hyperfine splitting, the value of which has been recently updated to  $\Delta_b^\mathrm{exp} = m_{\Upsilon(1S)} - m_{\eta_b} =  62.3 \pm 3.2 ~{\rm MeV}$~\cite{PDG2024}, after the reassessment of uncertainties in the old extractions of $m_{\Upsilon(1S)}$~\cite{Shamov:2022ajx}. Note that its value is about $2\sigma$ larger than predicted by a resummation of perturbative QCD series in Refs.~\cite{Kniehl:2003ap,Recksiegel:2003fm}.
Lattice QCD provides a way to compute that quantity model independently. While the initial lattice results pointed to a small value for $\Delta_b^\mathrm{latt} \lesssim 40$~MeV~\cite{CP-PACS:2000qyk,Liao:2001yh}, it subsequently became clear that working with much lighter sea quark masses lead to larger values for hyperfine splitting. Indeed, in the studies with $N_f = 2+1$ staggered light quarks and with the so called Fermilab approach to heavy quarks~\cite{El-Khadra:1996wdx}, $\Delta_b^\mathrm{latt} = 54 \pm 12$ MeV, was reported in Ref.~\cite{Burch:2009az}. Instead, by using the nonrelativistic QCD (NRQCD) treatment of heavy quark on the lattice, the value $\Delta_{b}^\mathrm{latt} \approx 60 \pm 8~{\rm MeV}$ and  $\Delta_b^\mathrm{latt} \approx 70 \pm 9$ MeV, has been obtained in~\cite{Meinel:2010pv} and~\cite{HPQCD:2011qwj}, respectively, thus consistent with $\Delta_b^\mathrm{exp}$.

Another recent experimental effort carried out by Belle focused on $\Upsilon(4S)\to h_b(1P)\gamma$ followed by $h_b(1P) \rightarrow \eta_b \gamma$~\cite{Belle:2015hnh}. That measurement yields $\Delta_b^\mathrm{exp}$ fully consistent with the value quoted above, and it allowed the authors to extract $\mathcal{B}(h_b\to \eta_b\gamma) = (56\pm 8\pm4)\%$, improving on their previous Ref.~\cite{Belle:2012fkf}. 
Besides the importance of confronting the theory with experiment, $\Delta_b$ and $\mathcal{B}(h_b\to \eta_b\gamma)$ are important for searches of physics beyond the Standard Model (BSM). For example, in the models in which the Higgs sector is extended by one or more Higgs doublets, one can check whether or not a CP-odd Higgs state ($A$) is light because in that case it can mix with a quarkonium with the same quantum number, such as $\eta_c$ or $\eta_b$. Mixing with $\eta_b$ has been abundantly studied in Refs.~\cite{Andreas:2010ms, Dermisek:2010mg, Domingo:2009tb, Dermisek:2006py}. Another BSM possibility is that of an axion-like particle $a$ (ALP), which arise in scenarios in which the mass ($m_a$) and the coupling ($f_a$) are independent and not related as $f_a m_a \approx f_\pi m_\pi$,  like in the usual QCD axion. ALP can therefore be heavy and, being a pseudoscalar, it can mix with $\eta_{c}$ or $\eta_{b}$, and alter the experimentally observed decay modes involving the pseudoscalar states identified as $\eta_{c}$ or $\eta_{b}$~\cite{Kim:1986ax,Merlo:2019anv,Bauer:2021mvw,DiLuzio:2024jip}. For that reason, it is very important to have a precise QCD based determination of the hyperfine splitting and of the relevant transition matrix elements. In this paper we focus on the latter.

Previous works in lattice QCD have shown that the radiative decays of charmonia, such as $J/\psi \rightarrow \eta_c \gamma$ and $h_c \rightarrow \eta_c \gamma$, can be computed with a satisfactory precision~\cite{Dudek:2006ej}, allowing for a systematic extrapolations to the continuum limit with a corresponding uncertainty under control~\cite{Becirevic:2012dc,Donald:2012ga,Colquhoun:2023zbc}. Following the methodology of Ref.~\cite{Becirevic:2012dc}, the present study seeks to improve upon these calculations by focusing on the electric dipole (E1) transition of charmonia, $h_{c}\to \eta_{c}\gamma$, which is then extended to the case of bottomia, $h_{b}\to \eta_{b}\gamma$, which is a new lattice result. For this calculation we employ five different fine lattice spacings in the range $a \in [0.049-0.09]~{\rm fm}$.

In contrast to the charmonium system, which can be directly simulated on current lattices, the bottomonium one is computationally challenging because the ultraviolet (UV) cut-off scale  employed in current simulations ($\propto 1/a$, the inverse lattice spacing) is smaller than the $b$-quark mass ($m_{b}$). To circumvent this problem, we use a sequence of heavy quark masses lighter than the physical $b$-quark, $m_{H} < m_{b}$, however larger than the physical $m_c$. In selecting the heavy quark masses we choose $a m_{H} \lesssim 0.5$, as to keep the ultraviolet (UV) cutoff effects under control.
In this respect, particularly beneficiary are the two ensembles of gauge field configurations, produced by the Extended Twisted Mass Collaboration (ETMC), with  lattice spacings $a \approx 0.058~\mathrm{fm}$ and $a \approx 0.049~\mathrm{fm}$, which allow us to reach the heavy quarks up to nearly three times the charm quark mass. To obtain the results at the physical point (corresponding to $m_{H} \rightarrow m_{b}$), we guided our extrapolation by phenomenological parameterizations inspired by NRQCD.

The remainder of this paper is organized as follows. In Section~\ref{sec:hc_etac}, we discuss the charmonium decay $h_{c} \rightarrow \eta_{c} \gamma$, where we work directly at the physical charm quark mass for which we provide a precise estimate of the relevant form factor and therefore the decay width. In Section~\ref{sec:hb_etab}, we focus on the bottomonium decay $h_{b} \to \eta_{b} \gamma$. We employ the ratio method of Ref.~\cite{ETM:2009sed} to reduce systematic errors, enhance statistical precision, and suppress lattice discretization effects in the determination of the radiative heavy onium decay width for $m_{c} < m_{H} < m_{b}$. We then present the results of the continuum limit extrapolation for each of the simulated heavy quark masses, discuss the various scaling relations used to extrapolate to the physical $b$-quark mass, and present our final results for the nonperturbative transition form factor describing $h_{b}\to\eta_{b}\gamma$. Finally, in Section~\ref{sec:comparison}, we compare our results to experimental measurements and/or existing theoretical calculations, and in Section~\ref{sec:conclusions} we summarize the implications of our results 
outlining potential directions for future research.

\section{Lattice Setup and Calculation of the $ h_c \to \eta_c \gamma $ Transition Form Factor}
\label{sec:hc_etac}

The transition matrix element describing the radiative decay $ h_c \to \eta_c \gamma $ is given by:
\begin{widetext}
\begin{align}
\label{eq:def_trans_ff}
\langle \eta_{c}(k) | J^{\mu}_{\rm em} | h_{c}(p,\varepsilon_{\lambda}) \rangle &=  2iQ_{c}\left\{ m_{h_{c}} F_{1}^{c}(q^{2})\left(\varepsilon^{\ast\mu}_{\lambda} - \frac{\varepsilon^{\ast}_{\lambda}\cdot q}{q^{2}}q^{\mu}\right) 
+ F_{2}^{c}(q^{2})(\varepsilon_{\lambda}^{\ast}\cdot q)\left[ 
\frac{m_{h_{c}}^{2} - m_{\eta_{c}}^{2}}{q^{2}}q^{\mu} -(p+k)^{\mu} \right]    \right\}\,,
\end{align}
\end{widetext}
where $J^{\mu}_{\rm em}$ is the electromagnetic current,
\begin{equation}
\label{eq:em_current}
    J^{\mu}_{\text{em}}(x) = \!\!\sum_{f=u,d,s,c} \!\! J_{f}^{\mu}(x) = \!\!\sum_{f=u,d,s,c} \!\! Q_f \, \bar{q}_f(x) \gamma^{\mu} q_f(x)\, ,
\end{equation}
and the sum runs over all quark flavors $ f=\{u,d,s,c\} $, $ q_f(x) $ is the quark field, and $ Q_f $ the quark electric charge in units of the elementary charge $ e $. In Eq.~(\ref{eq:def_trans_ff}), $ p $ and $ k $ are the momenta of the $ h_c $ and $ \eta_c $ mesons respectively, and $q=p-k$, while $ \varepsilon_\lambda $ is the polarization vector of the $ h_c $. The form factors at $q^2=0$ satisfy the relation $m_{h_c} F_1^c(0) = (m_{h_c}^2 -m_{\eta_c}^2)  F_2^c(0)$, and therefore the corresponding rate of the decay to a real photon can be expressed in terms of the single transition form factor $F_{1}^{c} \equiv F_{1}^{c}(0)$ through,
\begin{align}
\label{eq:decay_rate_charm}
\Gamma\left( h_{c} \to \eta_{c}\gamma \right) =  \frac{2Q_{c}^{2}}{3}\alpha_{\rm em}\frac{ m_{h_{c}}^{2} - m_{\eta_{c}}^{2}}{m_{h_{c}}}\,|F_{1}^{c}|^{2}~,
\end{align}
which thus encodes all the nonperturbative information on the decay. 

To compute the form factor $F^{c}_{1}$ on the lattice, we make use of the gauge configurations produced by the Extended Twisted Mass  Collaboration (ETMC) with $N_{f}=2+1+1$ dynamical Wilson-Clover twisted mass fermions. This framework guarantees the automatic $O(a)$ improvement of parity-even observables~\cite{Frezzotti:2003ni,Frezzotti:2004wz}. Basic information regarding the five lattice ensembles used in this work is provided in Table~\ref{tab:simudetails}, and further details can be found in  Ref.~\cite{ExtendedTwistedMassCollaborationETMC:2024xdf}. In this work we employ the mixed-action lattice setup introduced in~Ref.~\cite{Frezzotti:2004wz}, and described in the appendices of Ref.~\cite{ExtendedTwistedMassCollaborationETMC:2024xdf}. In this setup, the action of the valence quarks is discretized in the so-called Osterwalder--Seiler (OS) regularization, namely, 
\begin{widetext}
\begin{align}
\label{eq:tm_action}
S = \sum_{f=u,d,s,c}\sum_{x} \bar{q}_{f}(x) \left[ \gamma_{\mu}\bar{\nabla}_{\mu}[U] -ir_{f}\gamma^{5}(W^{\rm cl}[U] + m_{\rm cr}) + m_{f}  \right] q_{f}(x)~,
\end{align}
\end{widetext}
where $W^{\rm cl}[U]$ is the Wilson-Clover term~\cite{Sheikholeslami:1985ij}, $m_{\rm cr}$ is the critical mass, $m_{f}$ the quark mass of the flavor $f$ (with $m_{u}=m_{d}=m_{l}$), and $r_{f}=\pm 1$ is the sign of the twisted-Wilson parameter  for the flavor $f$ ($r_{u,c}= -r_{d,s}=1$). 
At each lattice spacing, the charm quark mass $m_{c}$ has been tuned to reproduce $m_{D_{s}}= 1967~{\rm MeV}$. 
\begin{table}[t]
\begin{ruledtabular}
\begin{tabular}{lccccc}
\textrm{ID} & $L/a$ & $a$ \textrm{fm} & $Z_{V}$ & $am_{c}$ & $N_{g}$ \\
\colrule
\textrm{A48} & $48$ & 0.0907(5) &  0.68700(15) & $0.2620$ & $300$ \\
\textrm{B64} & $64$ & 0.07948(11) &  0.706354(54) & $0.23157$  & $203$ \\
\textrm{C80} & $80$ & 0.06819(14)  & 0.725440(33) & $0.19840$  & $609$  \\
\textrm{D96} & $96$ & 0.056850(90) & 0.744132(31)   &  $0.16490$  &  $150$  \\
\textrm{E112} & $112$ & 0.04892(11) &  0.758238(18) & $0.14125$   & $93$
\end{tabular}
\end{ruledtabular}
\caption{\small\sl$N_{f}=2+1+1$ ETMC gauge ensembles used in this calculation. We give the spatial extent in lattice units $L/a$, the lattice spacing $a$, the bare charm quark mass $am_{c}$, the renormalization constant of the vector current $Z_{V}$ determined in Ref.~\cite{ExtendedTwistedMassCollaborationETMC:2024xdf} using the twisted-mass Ward identity, and the number of gauge configurations $N_{g}$. With the exception of the A48 ensemble which corresponds to a pion mass $m_{\pi}\simeq 175~{\rm MeV}$, all the ensembles have been generated at physical values of the light, strange and charm quark masses. \label{tab:simudetails}}
\end{table}

We work in the rest frame of the $h_{c}$ meson ($\mathbf{p}=0$) which means that to ensure $q^2=0$ one needs to give to the $\eta_{c}$ meson a three-momentum 
\begin{align}
\label{eq:momentum_etac}
|\mathbf{k}| = \frac{ m_{h_{c}}^{2} -m_{\eta_{c}}^{2} }{2m_{h_{c}}} \simeq 500~{\rm MeV}~. 
\end{align}
To estimate the form factor $F_{1}^{c}$ we consider the following three-point correlation function:
\begin{widetext}
\begin{align}
\label{eq:three-point}
 C_{\text{3pt}}^{\mu}(t_{h}; t_{J}) = \sum_{\mathbf{x}, \mathbf{y},\mathbf{z}} e^{i\mathbf{k}(\mathbf{x}-\mathbf{y})} \langle 0 | \mathcal{O}_{\eta_c}(\mathbf{x}, 0) J^{\mu}_{\rm em}(\mathbf{y}, -t_{J}) \mathcal{O}_{h_c}^\dagger(\mathbf{z}, -t_{h}) | 0 \rangle~,
\end{align}
\end{widetext}
where $\mathcal{O}_{\eta_{c}}$ and $\mathcal{O}_{h_{c}}$ are the (Gaussian-smeared) interpolating operators of the $\eta_{c}$ and $h_{c}$ mesons,
\begin{equation}
\label{eq:interpolators}
\begin{split}
\mathcal{O}_{\eta_{c}}(\mathbf{x},t) &= \sum_{\mathbf{y}}\bar{q}_{c}(\mathbf{x},t) G_{t}^{n}(\mathbf{x},\mathbf{y}) \gamma^{5} q_{c}(\mathbf{y},t) \,,  \\ 
\mathcal{O}_{h_{c}}(\mathbf{x},t) &= \sum_{\mathbf{y}}\bar{q}_{c}(\mathbf{x},t) G_{t}^{n}(\mathbf{x},\mathbf{y}) \sigma^{23} q_{c}(\mathbf{y},t)  \,,
\end{split}
\end{equation}
where $\sigma^{\mu\nu} = i[\gamma^{\mu},\gamma^{\nu}]/2$, and 
\begin{align}
\label{eq:G_def}
G_{t}(\mathbf{x},\mathbf{y}) = \frac{1}{1+ 6\kappa}\bigl( \delta_{\mathbf{x},\mathbf{y}} + \kappa H_{t}(\mathbf{x},\mathbf{y})    \bigr)\,,
\end{align}
with $H_{t}(\mathbf{x},\mathbf{y})$ being the Gaussian smearing operator
\begin{align}
H_{t}(\mathbf{x}, \mathbf{y}) = \sum_{\mu=1}^{3}\bigl[ U^{\star}_{\mu}(\mathbf{x},t)\delta_{\mathbf{x}+\hat{\mu},\mathbf{y}} + U^{\star\dagger}_{\mu}(\mathbf{x}-\hat{\mu},t)\delta_{\mathbf{x}-\hat{\mu},\mathbf{y}}    \bigr]\,,
\end{align}
where $U^{\star}_{\mu}(x)$ is the so-called APE-smeared links,  cf.~\cite{Becirevic:2012dc}.
We use the smearing parameter $\kappa=0.4$, and fix on each ensemble the number of steps $n$ in Eq.~\eqref{eq:interpolators} so as to obtain a smearing radius $r_{0}= a\sqrt{n}/\sqrt{\kappa^{-1}+6} \simeq 0.15~{\rm fm}$, for both $\eta_{c}$ and $h_{c}$~\cite{DiPalma:2024jsp}. The spatial momentum $\mathbf{k}$ is injected along the third spatial direction, i.e. $\mathbf{k} = (0,0,|\mathbf{k}|)$. With this choice, and with the interpolating operators~\eqref{eq:interpolators},  the form factor $F_{1}^{c}$ can be evaluated by computing only the component $C_{\rm 3pt}^{1}$ of the three-point correlation function~\eqref{eq:three-point}.

\begin{figure*}
\centering
\includegraphics[scale=0.44]{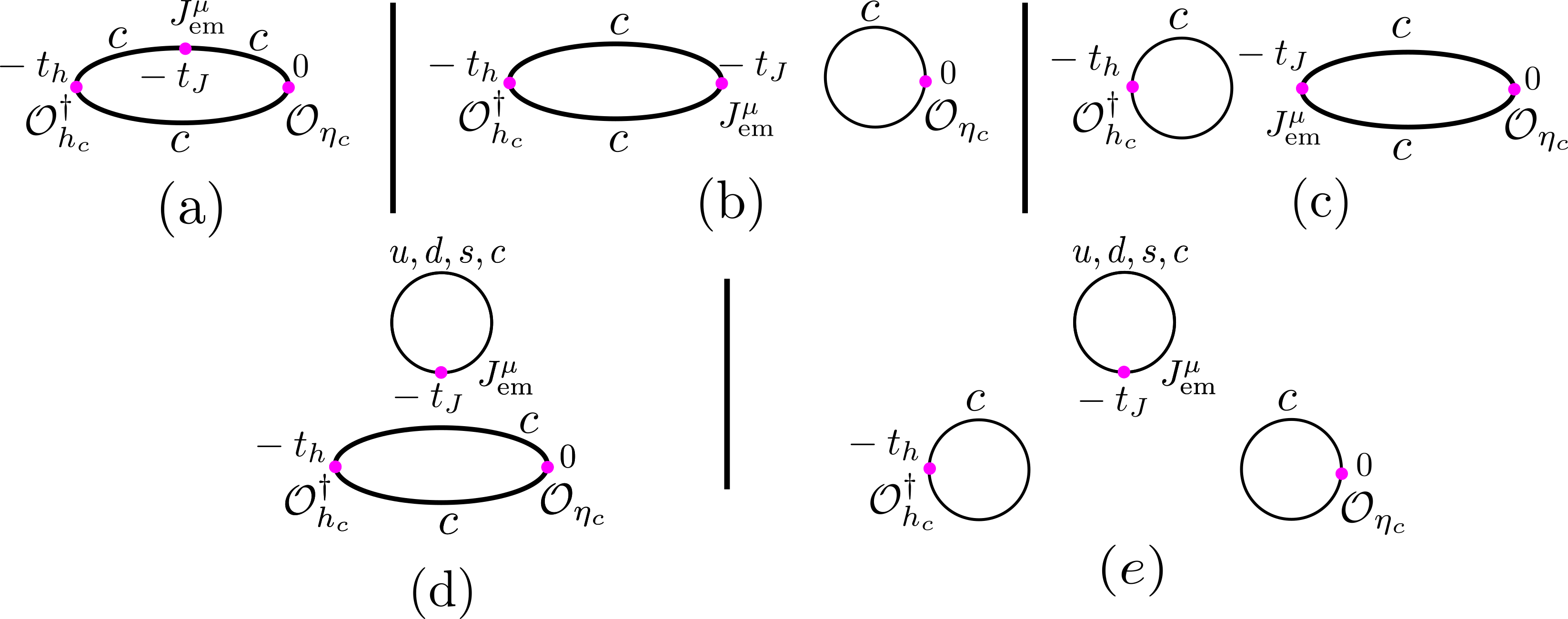}
\caption{\small\sl Wick contractions relevant to the correlation function~(\ref{eq:three-point}). The diagram (a) corresponds to the connected contribution, while (b), (c), (d) and (e) correspond to the disconnected ones. Diagrams (b), (c) and (e) are Zweig suppressed. For each diagram, we have indicated the flavor associated to each quark line. Light quark ($u$, $d$, $s$) contributions only appear in diagram (d) and diagram (e)  where the photon is emitted by a sea quark. These two diagrams vanish in the $\rm{SU}(3)$-limit $m_{u}=m_{d}=m_{s}$, while the effect of the $c$ sea quark should be negligible. \label{fig:conn-disc}}
\end{figure*}

The Wick contractions in Eq.~\eqref{eq:three-point} give rise to quark-connected and quark-disconnected contributions, cf. Fig.~(\ref{fig:conn-disc}). In the following, we will refer to these contributions as \textit{connected} and \textit{disconnected}, respectively.
The leftmost diagram corresponds to the dominant (connected) contribution, while (b), (c) and (d) correspond to the disconnected contributions. Note that the contributions corresponding to (b), (c) and (e) are expected to be very small due to the Zweig suppression.~\footnote{The Zweig suppression is expected to be even more effective for the $h_{b}\to \eta_{b}\gamma$ decay discussed in Sec.\ref{sec:hb_etab}.}  Instead, diagram (d) and diagram (e) respect the $\mathrm{SU}(3)$ suppression in which $u$, $d$ and $s$ contributions cancel when $m_{u}=m_{d}=m_{s}$. In this work we focus on the evaluation of the dominant connected diagram, leaving the evaluation of the disconnected contributions for future works.~\footnote{A lattice QCD study of Ref.~\cite{Hatton:2020qhk} show that the contributions arising from disconnected diagrams give a tiny contribution to the masses of charmonia.} Clearly, only the charm quark component $J_{c}^{\mu}$ of the electromagnetic current $J^{\mu}_{\rm em}$ contributes to the connected part of the correlation function $C_{\rm 3pt}^{\mu}(t_{h};t_{J})$. 

\begin{figure}
\centering
\includegraphics[scale=1]{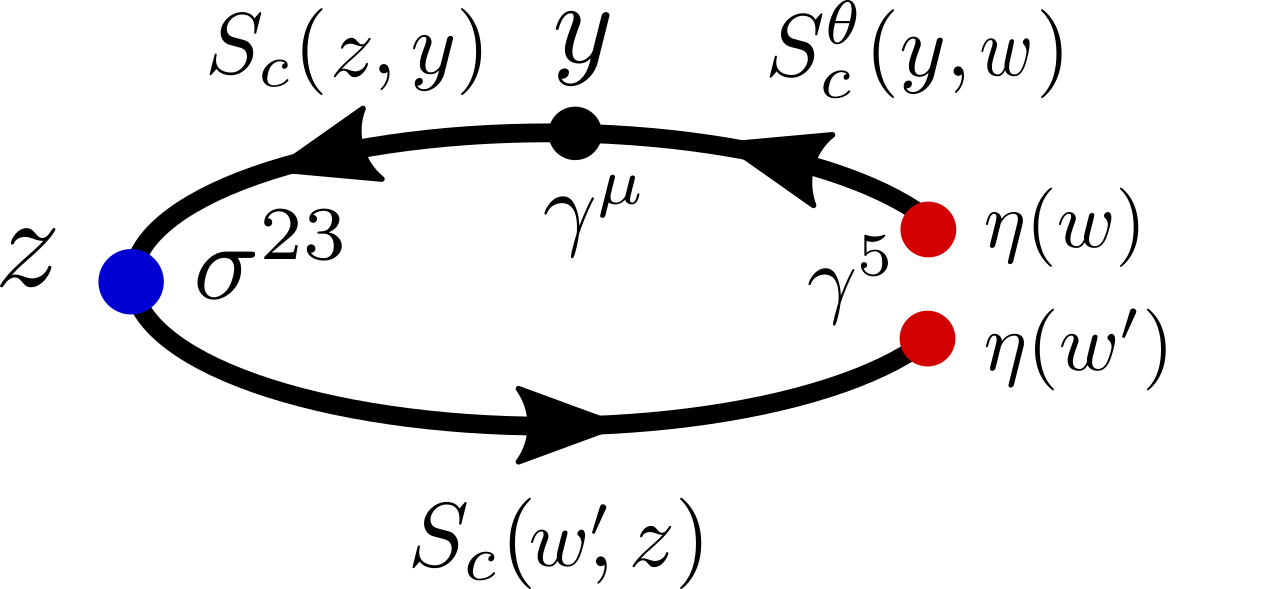}
\caption{\small\sl Schematic representation of the strategy adopted to evaluate the connected contribution to $C_{\rm 3pt}^{\mu}(t_{h}; t_{J})$. The sum over $\mathbf{x}$ is implemented stochastically by inverting the charm quark Dirac operator on a number $N_{\rm stoch}$ of the spatial stochastic sources $\eta$ (red vertices). The electromagnetic current is inserted in the upper quark line at the fixed time $-t_{J}$ (black vertex) and at all spatial positions $\mathbf{y}$. The blue vertex indicates $z=(\mathbf{z}, -t_{h})$, where the $h_{c}$ meson is created.      \label{fig:3pt_scheme}}
\end{figure}

As far as the connected contribution is concerned, we employ the twisted boundary conditions in order to tune the spatial momentum $\mathbf{k}$ to the value of Eq.~(\ref{eq:momentum_etac}). This is implemented by twisting the gauge links $U_{\mu}(x)$ on which one of the charm quark propagators is computed, namely 
\begin{align}
\label{eq:twisted_gauge}
U_\mu(x) \rightarrow U^\theta_\mu (x) = e^{i a\theta_{\mu} / L} U_\mu(x), \quad  \theta_\mu = (0, \vec{\theta})\,,
\end{align}
with the twisting-angle $\vec{\theta}$ set to
\begin{align}
\label{eq:theta}
\vec{\theta}=( 0, 0 , \theta_{z}^{c}) , \qquad \theta_{z}^{c} = \frac{L}{\pi}\,\frac{ m_{h_{c}}^{2} - m_{\eta_{c}}^{2}}{2m_{h_{c}}}~. 
\end{align}
The connected part of $C_{\rm 3pt}^{\mu}(t_{h}; t_{J})$ is evaluated as follows.~\footnote{To simplify the expressions we consider the case of local interpolating operators, i.e. $\kappa=0$.} By $S_{c}(x,y)$ and $S_{c}^{\theta}(x,y)$ we denote the charm quark propagators evaluated on the background field configurations corresponding to $U_{\mu}(x)$ and $U_{\mu}^{\theta}(x)$, respectively, so that $C_{\rm 3pt}^{\mu}(t_{h}; t_{J})$ can be written as:
\begin{widetext}
\begin{align}
\label{eq:3pt_explicit}
C_{\rm 3pt}^{\mu}(t_{h}, t_{J}) = 2\,\sum_{\mathbf{x},\mathbf{y},\mathbf{z}} \langle \tr \left[ S_{c}(x, z) \sigma^{23} S_{c}(z,y) \gamma^{\mu} S_{c}^{\theta}(y, x) \gamma^{5} \right]     \rangle\,,
\end{align}
\end{widetext}
where $x=(\mathbf{x},0)$, $y=(\mathbf{y}, -t_{J})$, $z=(\mathbf{z}, -t_{h})$. The trace $\tr\left[ \dots \right]$ is taken over the color and Dirac indices, and $\langle \dots \rangle$ indicates the average over the $\rm{SU}(3)$ gauge field configurations $U_{\mu}(x)$. The factor of two in Eq.~\eqref{eq:3pt_explicit} accounts for the contribution of the charge conjugated diagram in which the photon is emitted by the charm antiquark. The sum over $\mathbf{x}$ is evaluated stochastically by performing the inversion of the charm quark Dirac operator on a number $N_{\rm stoch}$ of spatial stochastic sources $\eta(w)$, placed at time $t=0$ (stochastic time wall sources), namely
\begin{equation}
\begin{split}
S_{c}^{\theta}(y,x) &\mapsto S_{c}^{\theta}(y,w)\eta(w)\,, \\[10pt]
S_{c}(z,x) &\mapsto S_{c}(z,w') \eta(w')\,,
\end{split}
\end{equation}
with $\eta(\mathbf{w}, w_{0}) = \delta_{w_{0}, 0} \bar{\eta}(\mathbf{w})$, and $\langle \bar{\eta}^{\dagger}(\mathbf{w'})\bar{\eta}(\mathbf{w})\rangle_{\eta} = \delta_{\mathbf{w'},\mathbf{w}}$. The backward propagator $S_{c}(x,z)$ is then obtained from $S_{c}(z,x)$ by using $\gamma^{5}$-hermiticity. Depending on the ensemble considered, we use up to $N_{\rm stoch}= 32$ spin-diluted stochastic sources per gauge configuration. A diagrammatic representation of the strategy adopted to evaluate the connected contribution to $C_{\rm 3pt}^{\mu}(t_{h}; t_{J})$ is shown in Fig.~(\ref{fig:3pt_scheme}).

We now discuss how the charm quark bilinears $\mathcal{O}_{\eta_{c}}$, $\mathcal{O}_{h_{c}}$, and the current $J_{c}^\mu$, are discretized on the lattice. When evaluating the connected diagrams using the twisted mass action, there are different options when it comes to discretizing the quark bilinears and they differ in the choice of the  Wilson parameter of the quark field, $r_{+}=\pm 1$, and of the antiquark field, $r_{-}= \pm 1$. They can be chosen to be opposite ($r_{+}=-r_{-}$) or equal ($r_{+} = r_{-}$). The two regularizations are known as the twisted mass (TM) and Osterwalder-Seiler regularizations, respectively. The results obtained using different combinations of TM and OS bilinears differ only by $O(a^{2})$ UV cutoff effects. A complete description of the two regularizations is given in Appendix~B of Ref.~\cite{ExtendedTwistedMassCollaborationETMC:2024xdf} to which we refer for further details. In this calculation we opt to use the OS regularization for the electromagnetic current, while $\mathcal{O}_{\eta_{c}}$ and $\mathcal{O}_{h_{c}}$ are regularized as TM bilinears.\footnote{ This choice for the interpolating operator of the $h_{c}$ meson, avoids the mixing with $1^{--}$ ($J/\Psi$) and $1^{++}$ ($\chi_{c1}$) states. This is a consequence of the exact $P \times G$ symmetry, where $P$ is parity and $G$ is $G-$parity, enjoyed by the TM regularization (see also Ref.~\cite{Petry:2008rt}).} With this choice, the interpolating operators are given by
\begin{equation}
\label{eq:interpolators_TM}
\begin{split}
\mathcal{O}_{\eta_{c}}(\mathbf{x},t) &= \sum_{\mathbf{y}}\bar{q}_{c}(\mathbf{x},t) G_{t}^{n}(\mathbf{x},\mathbf{y}) \gamma^{5} q'_{c}(\mathbf{y},t)\,,  \\ 
\mathcal{O}_{h_{c}}(\mathbf{x},t) &= \sum_{\mathbf{y}}\bar{q}_{c}(\mathbf{x},t) G_{t}^{n}(\mathbf{x},\mathbf{y}) \sigma^{23} q'_{c}(\mathbf{y},t) \,,
\end{split}
\end{equation}
and the quark action of the field $q'_{c}(x)$ is equal to that of the $q_{c}(x)$ field in Eq.~(\ref{eq:tm_action}) after replacing $r'_{c} = -r_{c}$. The charm quark component of the electromagnetic current, instead, is regularized as $J_{c}^{\mu}(x)= \bar{q}_{c}(x)\gamma^{\mu}q_{c}(x)$ with $r_{c}=1$.
With this choice, the renormalization constant of the local electromagnetic current is given by $Z_{V}(g_{0}^{2})$, the value of which is given in  Tab.~\ref{tab:simudetails} for each lattice ensemble used in this work. If we use $S_{c}^{r}$ and $S_{c}^{\theta, r}$ to denote the charm quark propagators corresponding to a given value of $r= \pm 1$, then Eq.~\eqref{eq:3pt_explicit} gets modified by the replacements, 
$S_{c}^{\theta}(y,x) \to S_{c}^{\theta, +}(y,x)$, and  
$S_{c}(z,y) \to S_{c}^{+}(z,y)$ and $S_{c}(x,z)  \to S_{c}^{-}(x,z)$. Note that  the $\gamma^{5}$-hermiticity of the twisted Wilson quark propagator reads:
\begin{align}
S^{\pm \dagger}_{c}(x,y) = \gamma^{5} S^{\mp}(y,x) \gamma^{5} \,.
\end{align}

\subsection{Extraction of the form factor $F_{1}^{c}$}
\label{sec:extr_form_factor_charm}
We first need to consider the two point functions of interpolating fields $\mathcal{O}_{\eta_{c}}$ and $\mathcal{O}_{h_{c}}$, respectively, in order to extract their mass/energy and their couplings. We define,
\begin{equation}
\label{eq:2pt_def}
\begin{split}
    C_{\text{2pt}}^{\eta_c}(t) &= \sum_{\mathbf{x}} e^{i\mathbf{k}\mathbf{x}} \langle 0 | \mathcal{O}_{\eta_c}(\mathbf{x}, t) \mathcal{O}_{\eta_c}^\dagger(\mathbf{0}, 0) | 0 \rangle\,, \\
    C_{\text{2pt}}^{h_c}(t) &= \sum_{\mathbf{x}} \langle 0 | \mathcal{O}_{h_c}(\mathbf{x}, t) \mathcal{O}_{h_c}^\dagger(\mathbf{0}, 0) | 0 \rangle\,.
\end{split}
\end{equation}
and again neglect the Zweig suppressed disconnected contributions to $C_{\rm 2pt}^{\eta_{c}}(t)$ and $C_{\rm 2pt}^{h_{c}}(t)$. 
In the limit of large Euclidean time $ t $, the two point correlation functions behave as:
\begin{equation}
\label{eq:2pt_asympt}
\begin{split}
    C_{\text{2pt}}^{\eta_c}(t) &= \frac{|Z_{\eta_c}|^2}{2 E_{\eta_c}} \left( e^{-E_{\eta_c} t} + e^{-E_{\eta_c} (T-t)}   \right) + \ldots ,  \\
     C_{\text{2pt}}^{h_c}(t) &= \frac{|Z_{h_c}|^2}{2 m_{h_c}} \left( e^{-m_{h_c} t} + e^{-m_{h_c} (T-t)}  \right ) + \ldots~,
\end{split}
\end{equation}
where $ m_{h_c} $ and $ E_{\eta_c} $ are the mass and energy (due to the nonzero momentum $\mathbf{k}$) of the corresponding meson, and 
\begin{align}
Z_{\eta_{c}} = \langle 0 | \mathcal{O}_{\eta_{c}} | \eta_{c}(k) \rangle\,, \qquad 
Z_{h_{c}} = \langle h_{c}(p,\epsilon_{\lambda}) | \mathcal{O}^{\dagger}_{h_{c}} | 0 \rangle\,.
\end{align}
The ellipses in Eq.~\eqref{eq:2pt_asympt} denote contributions that vanish in the large time limit, $0 \ll t \ll T$. We also computed $C_{\text{2pt}}^{\eta_c}(t)$ with $\mathbf{k}=\bf{0}$, which is needed to extract $m_{\eta_{c}}$. 

The interpolating field $\mathcal{O}^{\dag}_{h_{c}}$ in Eq.~(\ref{eq:interpolators}) creates an $h_{c}$ meson with polarization $\varepsilon^{\mu}_{\lambda} = -\delta^{\mu 1}$. 
Since we chose $q^{1}=p^{1}=0$, it is then straightforward to use Eq.~(\ref{eq:def_trans_ff}) and combine the above mentioned quantities into 
\begin{align}
\label{eq:FF_estimator}
\overline{F}_{1}^{c}(t_{h}; t_{J}) =\frac{i Z_{V}(g_0^2)}{2Q_{c}} \frac{ 4  E_{\eta_{c}}}{Z_{\eta_{c}} Z_{h_{c}}}  \  e^{E_{\eta_{c}}t_{J}} \  e^{m_{h_{c}}(t_{h}-t_{J})} \ C_{\rm 3pt}^{1}(t_{h};t_{J})\,,
\end{align}
which, for large Euclidean time separations, gives the desired form factor $F_{1}^{c}\equiv F_{1}^{c}(0)$:
\begin{align}
\overline{F}_{1}^{c}(t_{h}; t_{J}) \xrightarrow{\begin{array}{c}
\scriptstyle t_{J} \to \infty \\
\scriptstyle t_{h} - t_{J} \to \infty
\end{array}}
            F_{1}^{c}\,. 
\end{align}
Note again that we included $Z_{V}(g_0^2)$ in Eq.~\eqref{eq:FF_estimator}, the values of which are listed in Table~\ref{tab:simudetails}.

As already discussed in the previous Section, and illustrated in Fig.~\ref{fig:3pt_scheme}, we compute $C_{\rm 3pt}^{\mu}(t_{h};t_{J})$ by placing the interpolating field of the $\eta_{c}$ meson at a fixed time $t=0$, the charm quark component of the electromagnetic current is then inserted in a charm quark line at a fixed $-t_{J} < 0$, while the time $-t_{h} \ll -t_{J}$, at which the $h_{c}$ is created, corresponds to the \textit{sink} of the correlation function, which is thus known for all times $-t_{h}$. This allows us to carefully monitor the onset of the dominance of the $h_{c}$ state. Since in our computational setup the time $t_{J}$ is fixed, we must ensure to choose a sufficiently large $t_{J}$ to be able to isolate the $\eta_{c}$ meson state. The advantage of placing the interpolator corresponding to $\eta_{c}$ at the source of the correlation function is that $t_{J}$ can be chosen reasonably large without a significant loss of the signal, since the signal-to-noise ratio (S/N) of $C_{\rm 3pt}^{\mu}(t_{h};t_{J})$ practically only depends on the time distance $| t_{J}- t_{h} |$.~\footnote{ For a fixed $t_{J}-t_{h}$, according to the Parisi-Lepage theorem, the S/N ratio of $C_{\rm 3pt}^{\mu}(t_{h};t_{J})$ scales as $e^{-(E_{\eta_{c}}-m_{\eta_{c}})t_{J}} \simeq e^{-|\mathbf{k}|^{2} t_{J} / 2m_{\eta_{c}}}$,  with $|\mathbf{k}|^{2} / 2m_{\eta_{c}} \simeq 40~{\rm MeV}$. The S/N ratio thus decreases very slowly as $t_{J}$ increases. In the bottomonium case, to be discussed in the next Section, the deterioration of the signal
is even milder because the momentum transfer $|\mathbf{k}|$ is still of $O(500~{\rm MeV})$ but $m_{\eta_{b}} \simeq 9.4
~{\rm GeV}$.} As it will be detailed in the next subsection, we have considered two different values for $t_{J} \simeq 1.6, 2.4~{\rm fm}$, and checked that no difference in the results for $\overline{F}^{c}_{1}(t_{h};t_{J})$ are visible within statistical uncertainties.

\subsection{Numerical results}
\label{sec:num_results_hc}
\begin{figure}[]
\centering
\includegraphics[scale=0.35]{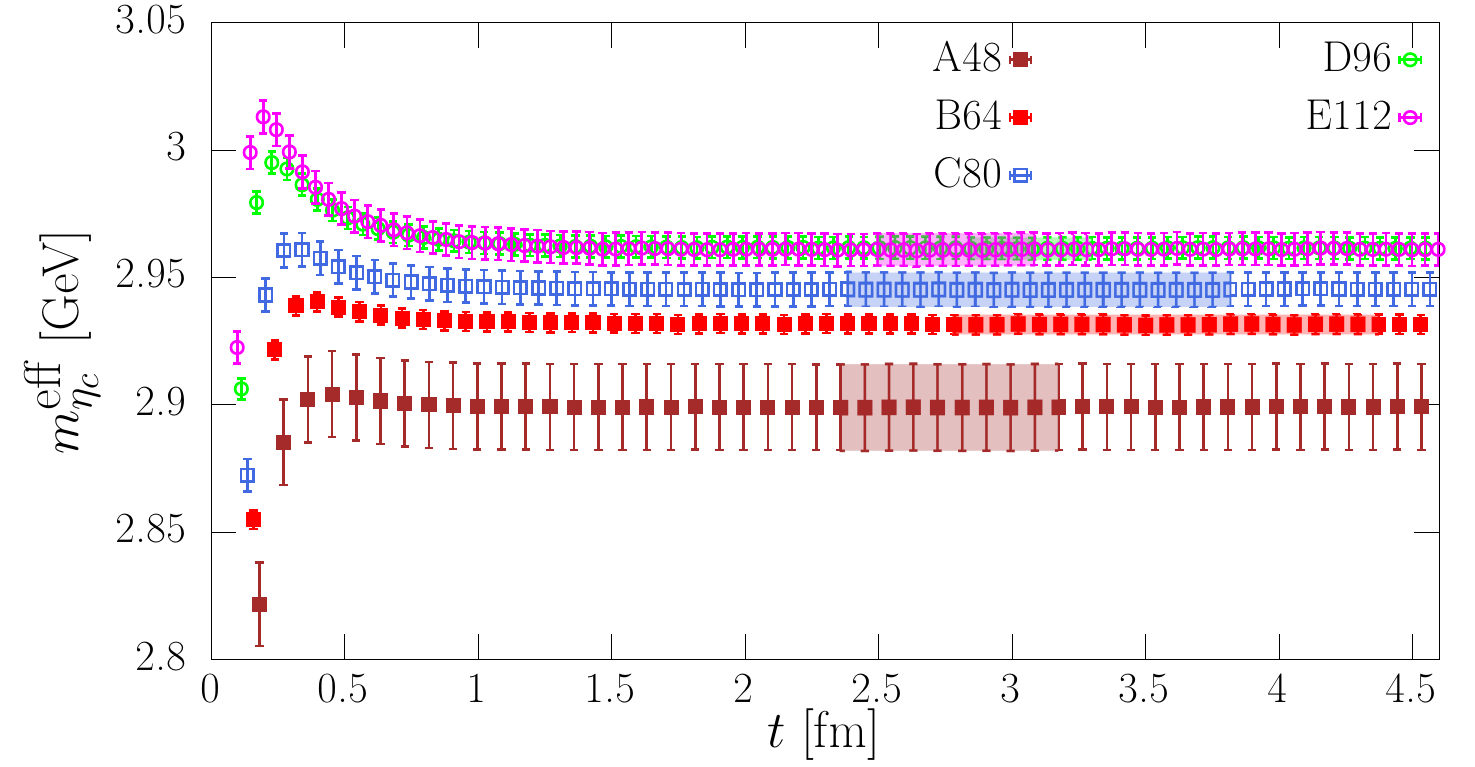}
\includegraphics[scale=0.35]{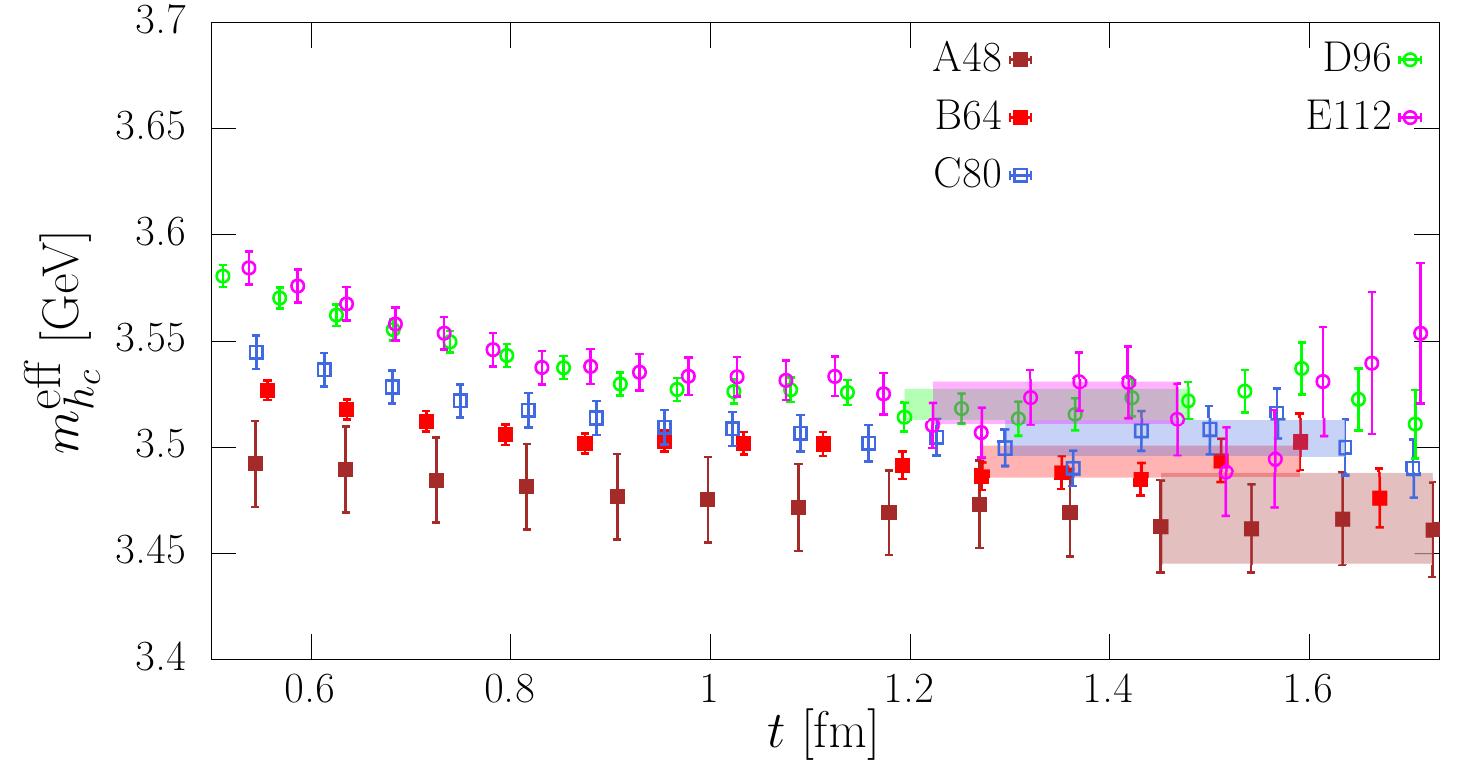}
\caption{\small\sl Effective mass of the $\eta_{c}$ meson (left) and of the $h_{c}$ meson (right) for all four gauge ensembles used for the present calculation. The coloured bands correspond to our estimate of the $\eta_{c}$ and $h_{c}$ masses on each ensemble. \label{fig:hc_etac_masses}}
\end{figure}

\begin{figure*}
\centering
\includegraphics[scale=0.35]{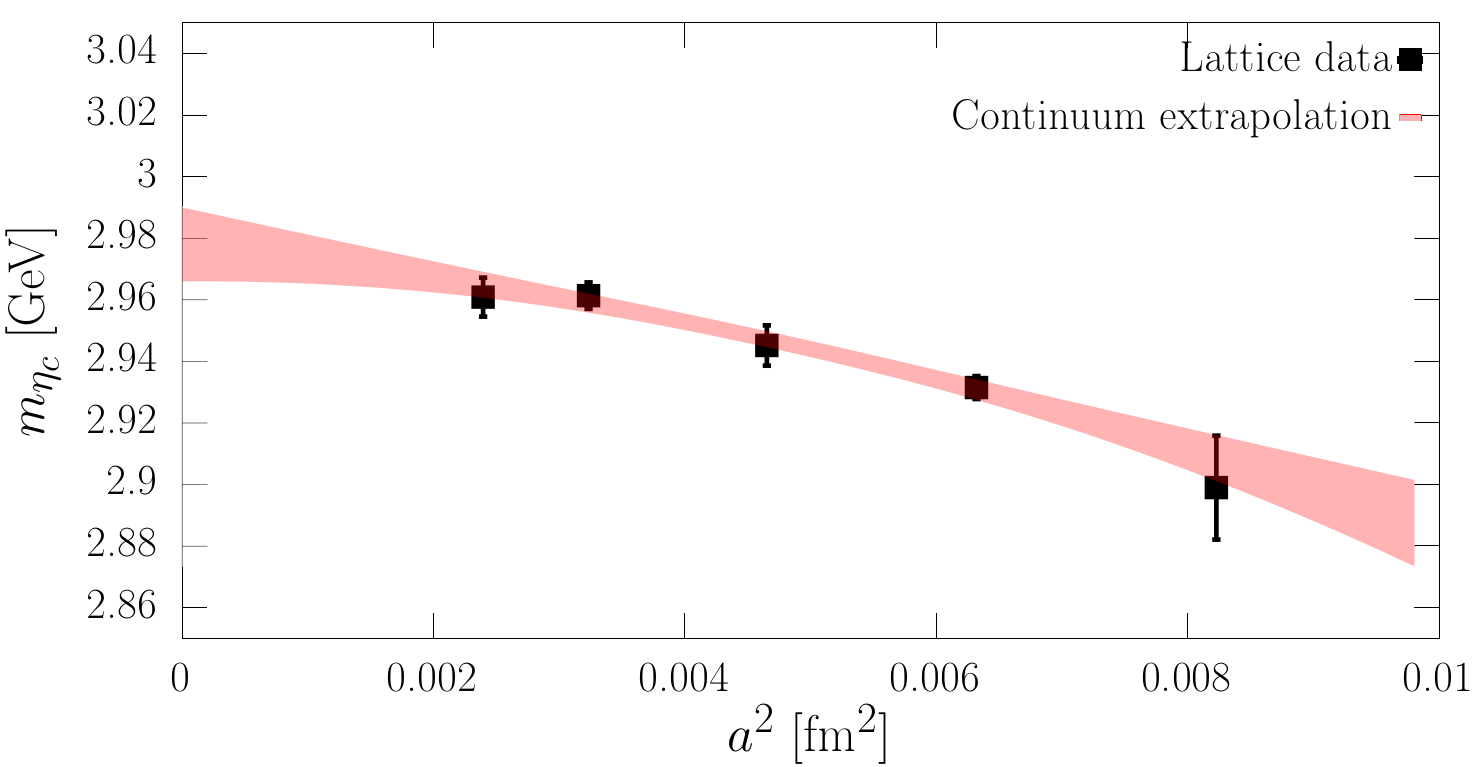}
\includegraphics[scale=0.35]{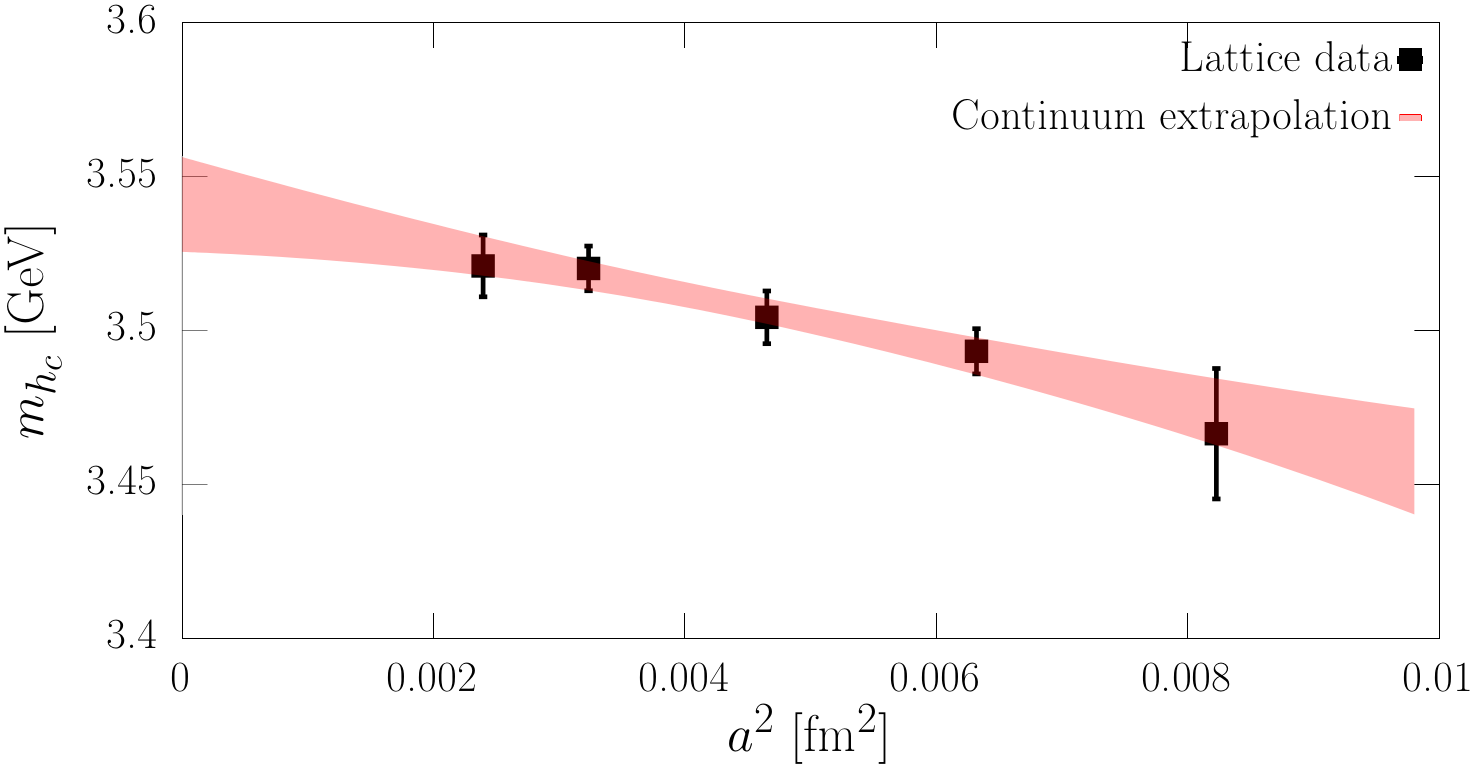}
\caption{\small\sl Extrapolation of $\eta_{c}$ (left) and $h_{c}$ (right) masses to the continuum limit ($a\to 0$). The colored bands correspond to the best fit obtained after performing linear and quadratic fits in $a^{2}$, and then combining them using the BAIC. The reduced $\chi^{2}$ of all the fits are always below one. \label{fig:hc_etac_extr}}
\end{figure*}

On all the ETMC gauge ensembles listed in Tab.~\ref{tab:simudetails} we computed the correlation functions $C_{\rm 2pt}^{h_{c}}(t), C_{\rm 2pt}^{\eta_{c}}(t)$ and $C_{\rm 3pt}^{\mu}(t_{h}; t_{J})$. In Fig.~\ref{fig:hc_etac_masses} we show the effective mass plots for $\eta_{c}$ and $h_{c}$ mesons, for all the lattice spacings. 
Note that since the charm quark mass has been fixed by $m_{D_{s}} =1967~\mathrm{MeV}$ at each value of the lattice spacing, due to the finite lattice artefacts the values of $m_{\eta_{c}}$ and $m_{h_{c}}$ are not the same for all ensembles. Owing to the fact that $\eta_{c}$ is the lightest charmonium state the S/N ratio of $C_{\rm 2pt}^{\eta_{c}}(t)$ does not decrease with $t$, as it is clear from Fig.~\ref{fig:hc_etac_masses}. This is in contrast to the case of $C_{\rm 2pt}^{h_{c}}(t)$, for which the effective mass becomes exponentially noisier as the time $t$ increases. 
To estimate the systematic error on $ m_{h_c} $, we additionally evaluated the smeared-local correlation function of the $ h_c $ meson, which is obtained by replacing one Gaussian-smeared operator $\mathcal{O}^\dagger_{h_c}$ in Eq.~\eqref{eq:interpolators} by the local one ($\kappa=0$). The difference between the effective mass plateaus obtained by considering the smeared-local and smeared-smeared correlation functions has been included in systematic uncertainty. The colored bands in Fig.~\ref{fig:hc_etac_masses} already account for that systematics. Note that the larger error for the A48 ensemble is due to uncertainty on the lattice spacing of that ensemble ($\simeq 0.6\%$). The extrapolation to the continuum limit of the lattice results for $m_{\eta_{c}}$ and $m_{h_{c}}$ is shown in Fig.~\ref{fig:hc_etac_extr}. 
The continuum fits are performed both linearly and quadratically in $a^{2}$. These two fits are then combined via the Bayesian Akaike Information Criterion (BAIC)~\cite{Neil:2022joj}, the method which we now briefly summarize.  Let $x_{1}, \ldots ,x_{N}$ be the outcomes of N different fits. The final central value and the total error are given by
\begin{equation}
\label{eq:BMA_ave}
\bar{x} = \sum_{n=1}^{N}\, w_{n}\,x_{n}\,,\qquad \sigma^{2} = \sigma_{\mathrm{stat}}^{2} + \sum_{n=1}^{N}\, w_{n}(x_{n} - \bar{x})^{2}\,,
\end{equation}
where $\sigma_{\mathrm{stat}}$ is the statistical error of $\bar{x}$, and $w_{1},\ldots, w_{N}$ are the weights normalized to one and given by
\begin{equation}
\label{eq:BAIC_weight}
w_{n} \propto \exp\Bigl[-\bigl(\chi_{n}^{2} + 2\,N_{n}^{\mathrm{pars}} - 2\,N_{n}^{\mathrm{data}}\bigr)\Bigr]~,
\end{equation}
with  $\chi_{n}^{2}$ being the chi-squared of the $n$-th fit, with its corresponding number of free parameters ($N_{n}^{\mathrm{pars}}$) and of the data points ($N_{n}^{\mathrm{data}}$). 

In the continuum limit we obtain
\begin{align}
\label{eq:etac_hc_cont_val}
m_{\eta_{c}} &= 2.978(12)~\mathrm{GeV}\,, \qquad m_{h_{c}} = 3.541(15)~\mathrm{GeV}\,,
\end{align}
which agree very well with the current experimental values, $m_{\eta_{c}}^\mathrm{exp} = 2.9841(4)~\mathrm{GeV}$ and $m_{h_{c}}^\mathrm{exp} = 3.52537(14)~\mathrm{GeV}$~\cite{PDG2024}. These results also suggest that the impact of neglecting the annihilation diagrams in $C_{\rm 2pt}^{\eta_{c}}$ and in $C_{\rm 2pt}^{h_{c}}(t)$, is small, i.e. of the order of our statistical uncertainty which is about $0.4\%$ for both $m_{\eta_{c}}$ and  $m_{h_{c}}$. This is also in line with Ref.~\cite{Hatton:2020qhk} where the contribution of the disconnected diagram to $m_{\eta_{c}}$ was found to be about $+7(1)~{\rm MeV}$.

We now turn to our determination of $ F_{1}^{c} $, starting with a discussion on the tuning of the momentum $\mathbf{k}$ of the $\eta_c$ at nonzero lattice spacing. The relation in Eq.~(\ref{eq:momentum_etac}) can be used by either plugging $ m_{\eta_c} $ and $ m_{h_c} $ obtained at each lattice spacing, or by using the experimental results, which, as we have just shown, agree well with our continuum extrapolated results in Eq.~(\ref{eq:etac_hc_cont_val}). These two choices lead to determinations of $ F_{1}^{c} $ that differ only by $O(a^2)$ UV cutoff effects. We opt for $ m_{\eta_c}^\mathrm{exp} $ and $ m_{h_c}^\mathrm{exp} $ and tune the momentum $\mathbf{k}$ for each lattice spacing.

\begin{figure}
\centering
\includegraphics[scale=0.49]{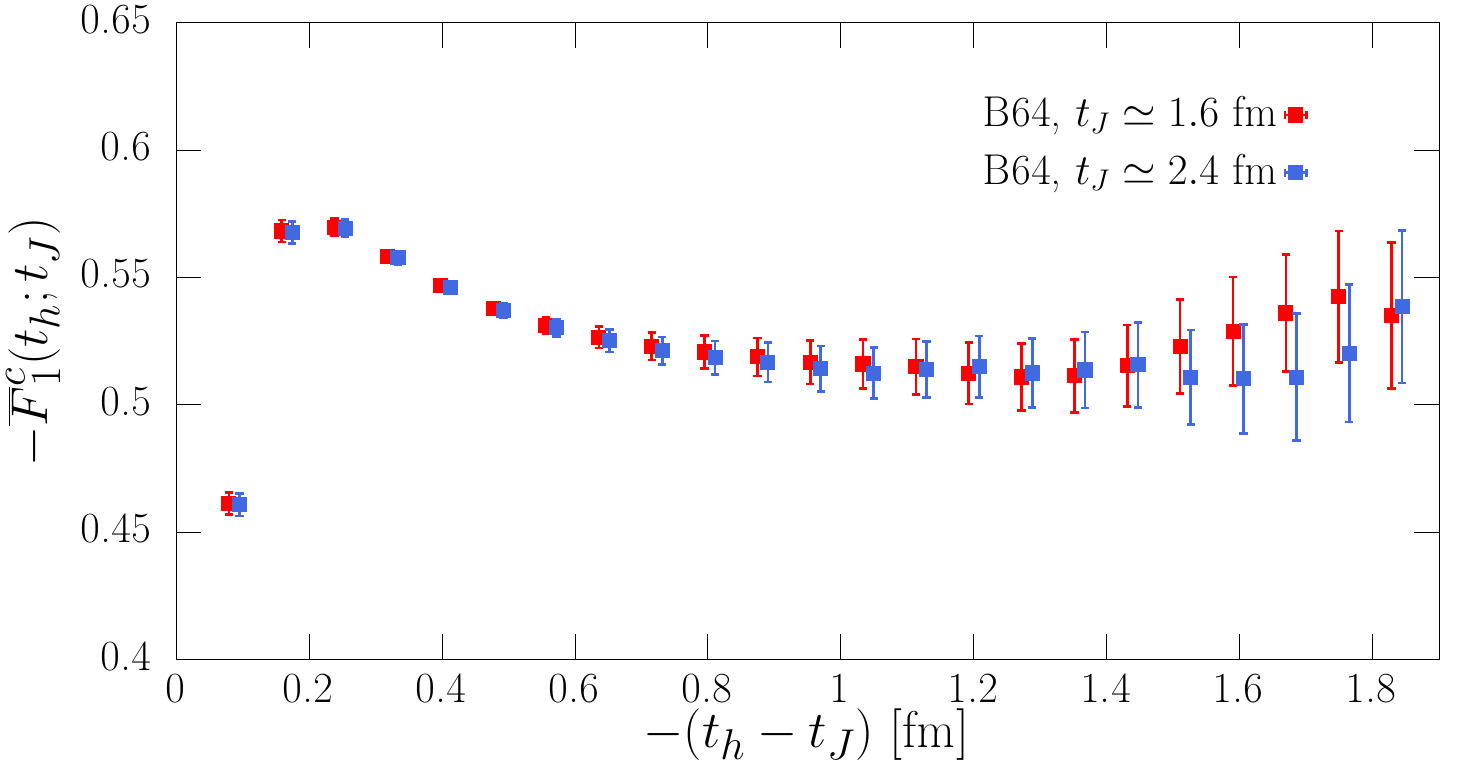}
\caption{\small\sl Comparison between $\overline{F}_{1}^{c}(t_{h}, t_{J})$ with $t_{J} \simeq 1.6~\mathrm{fm}$ (red) and with $t_{J} \simeq 2.4~\mathrm{fm}$ (blue) on the B64 ensemble. The datapoints at $t_{J}\simeq 2.4~\mathrm{fm}$ have been slightly shifted horizontally for easier comparison. \label{fig:3pt_test_B64}}
\end{figure}

As already anticipated, in our calculation of the correlation function $C_{\rm 3pt}^{\mu}(t_{h}; t_{J})$, the time $-t_{J}$ of the transition operator is always fixed, while $t_{h}$ runs over the lattice. It is thus important to choose sufficiently large $t_{J}$ in order to ensure the dominance of $\eta_{c}$.  We carried out a test on the B64 ensemble, and computed $C_{\rm 3pt}^{\mu}(t_{h}; t_{J})$ for two different values of $t_{J} \sim 1.6, 2.4~{\rm fm}$. In Fig.~\ref{fig:3pt_test_B64} we compare the corresponding $\overline{F}^{c}_{1}(t_{h}; t_{J})$, cf. Eq.~\eqref{eq:FF_estimator}.
Clearly, the two results are in excellent agreement. The difference in statistical uncertainties between the two cases is consistent with expectations of S/N of $ C_{\rm 3pt}^{\mu}(t_{h}, t_{J}) $ as a function of $ t_{J} $ (with $ t_{J}-t_{h} $ fixed), as discussed at the end of the previous subsection. On the basis of these observations we decided to fix $t_{J} \simeq 1.7~\mathrm{fm}$ for all other ensembles. 

\begin{figure}[t]
\centering
\includegraphics[scale=0.42]{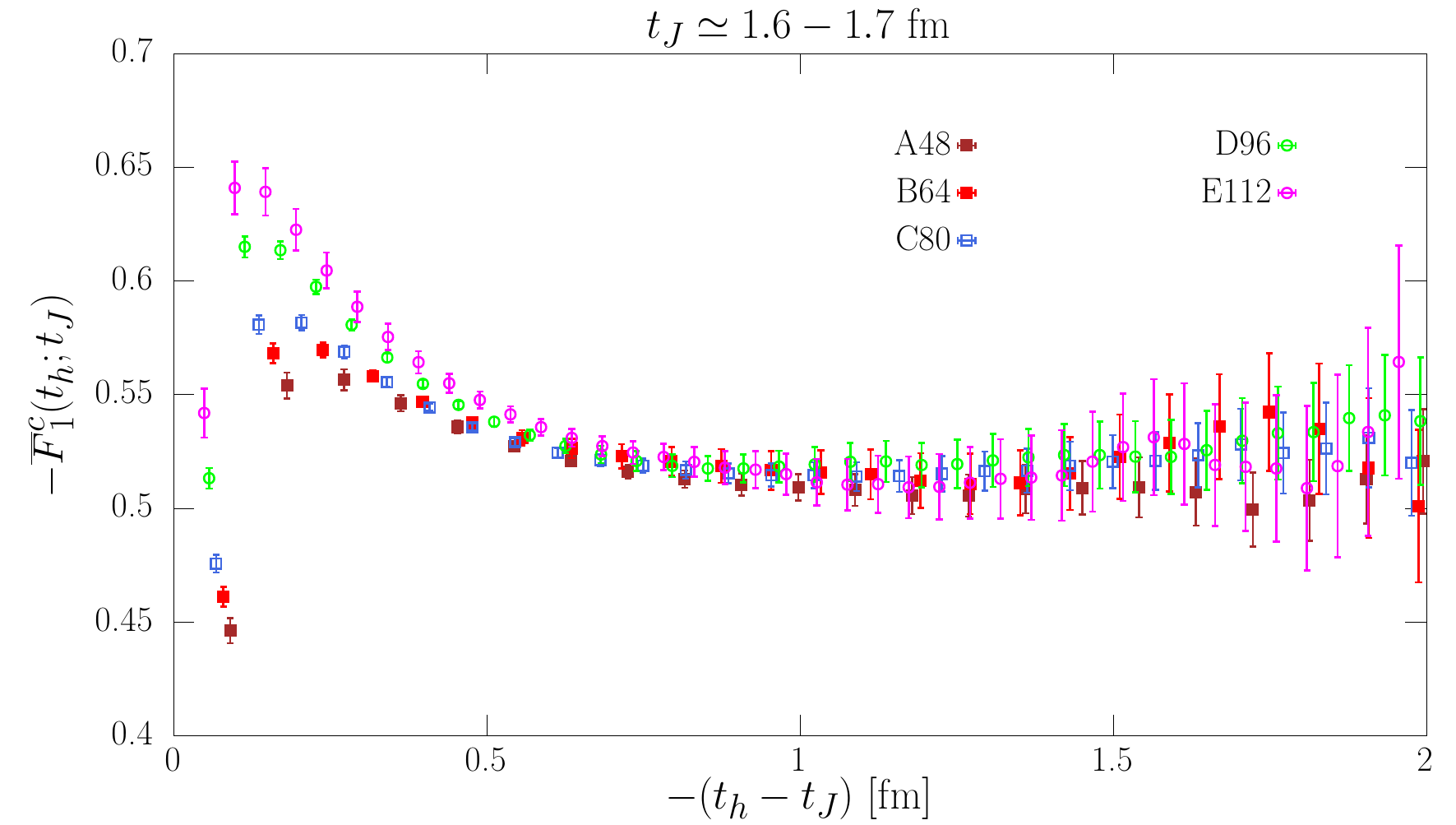}
\caption{\small\sl The function $\overline{F}_{1}^{c}(t_{h}; t_{J})$, defined in Eq.~\eqref{eq:FF_estimator}, as obtained for all lattice ensembles, cf. Tab.~\ref{tab:simudetails}, and for $t_{J}$ fixed as discussed in the text.   \label{fig:F1_all_beta}}
\end{figure}

In Fig.~\ref{fig:F1_all_beta} we show $\overline{F}_{1}^{c}(t_{h}; t_{J})$ for all our lattices. Quite remarkably, the UV cutoff effects are very small, and the results at all lattice spacings are in agreement within statistical errors. That feature has already been observed in Ref.~\cite{Becirevic:2012dc} in which the twisted mass Wilson quarks were used but without the Clover term. We then extract the form factor $F_{1}^{c}$ by fitting $\overline{F}_{1}^{c}(t_{h}; t_{J})$ to a constant in the time interval $ t_{J}-t_{h} \in (0.9, 1.3)~\mathrm{fm}$.

\begin{table}[t]
\begin{ruledtabular}
\begin{tabular}{lccccc}
$a~[{\rm fm}]$ & $0.0907\,(5)$ & $0.07948\, (11)$  & $0.06819\,(14)$    &  $0.056850\,(90)$ &  $0.04892\,(11)$   \\
\colrule
$-F_{1}^{c}$  & $0.5077(74)$ & $0.514(12)$ & $0.5148(64)$ &  $0.5197(88)$ & $0.512(12)$  
\end{tabular}
\end{ruledtabular}
\caption{\small\sl Values of the form factor $F_{1}^{c}$ for each gauge ensemble employed for the present calculation. \label{tab:F1c}}
\end{table}

\begin{figure*}[!t]
\centering
\includegraphics[scale=0.5]{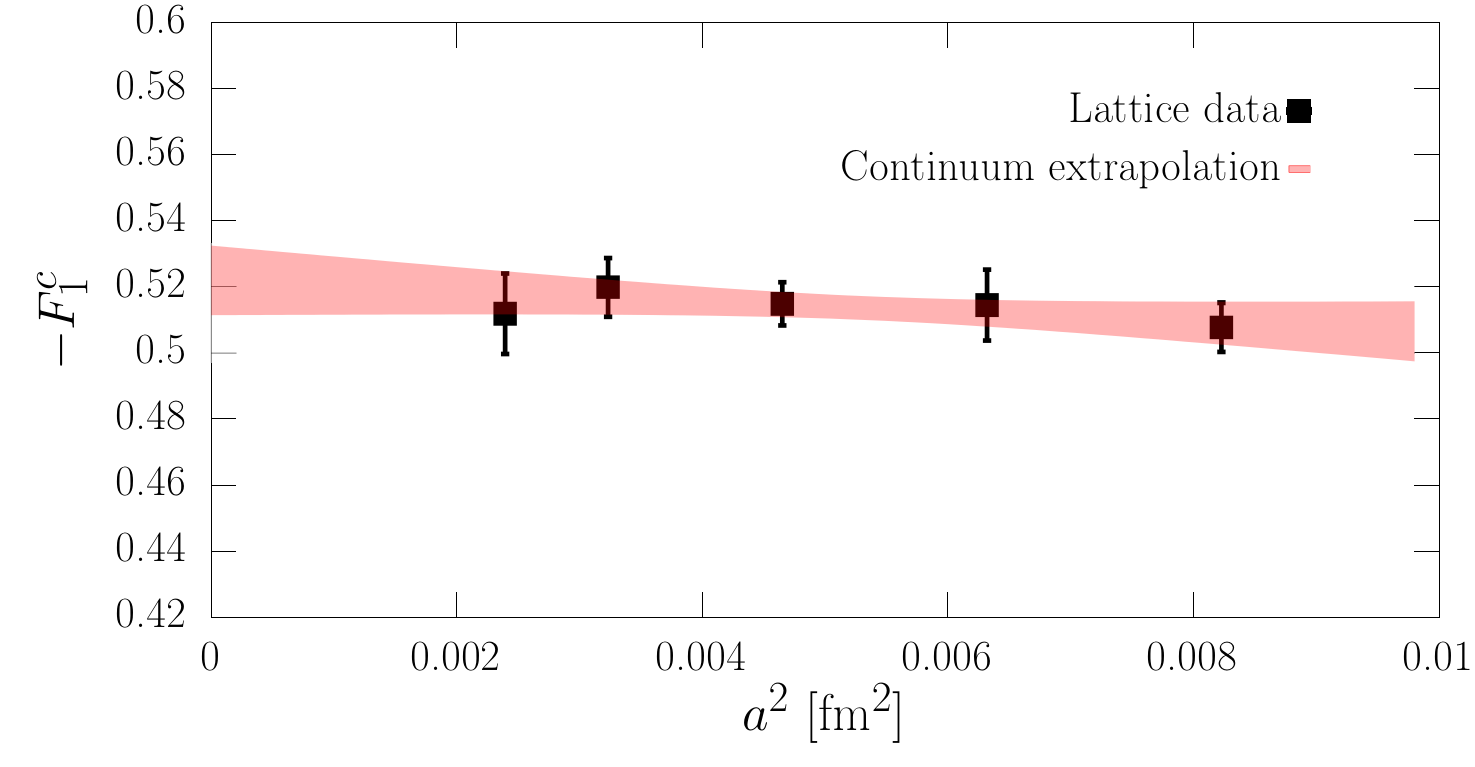}
\caption{\small\sl Continuum-limit extrapolation of the form factor $F_{1}^{c}$. The coloured band corresponds to the best-fit function obtained after performing a linear fit in $a^2$ to our lattice data. The reduced $\chi^{2}$ of the fit is around $0.2$.    \label{fig:cont_extr_F1c}}
\end{figure*}

The results of $F_{1}^{c}$ are presented in Tab.~\ref{tab:F1c}. We then extrapolate these results to the continuum limit by a simple linear $a^{2}$ fit, cf. Fig.~\ref{fig:cont_extr_F1c}. The corresponding $\chi^{2}/\mathrm{dof} \simeq 0.2$ is very good and our final result for the form factor describing $h_{c} \to \eta_{c} \gamma$ is
\begin{align}\label{eq:F1c}
\boxed{~F_{1}^{c} = -0.522(10)\,.~}
\end{align}
The uncertainty is more than twice smaller than the one reported in Ref.~\cite{Becirevic:2012dc}. In Sec.~\ref{sec:comparison}, we will compare our result for $F_{1}^{c}$ with existing theoretical determinations and with the available experimental results. We now turn to the analysis of the $h_{b} \to \eta_{b}\gamma$ decay.

\section{Calculation of the $h_{b}\to \eta_{b}\gamma$ transition form factor}
\label{sec:hb_etab}
Analogously to Eq.~(\ref{eq:def_trans_ff}) the transition matrix element describing $h_{b}\to \eta_{b}\gamma^\ast$ is decomposed as
\begin{align}
\label{eq:def_trans_ffb}
\langle \eta_{b}(k) | J^{\mu}_{\rm em} | h_{b}(p,\varepsilon_{\lambda}) \rangle &=  2iQ_{b}\left\{ m_{h_{b}} F_{1}^{b}(q^{2})\left(\epsilon^{\ast\mu}_{\lambda} - \frac{\epsilon^{\ast}_{\lambda}\cdot q}{q^{2}}q^{\mu}\right) 
+ F_{2}^{b}(q^{2})(\epsilon_{\lambda}^{\ast}\cdot q)\left[ 
\frac{m_{h_{b}}^{2} - m_{\eta_{b}}^{2}}{q^{2}}q^{\mu} -(p+k)^{\mu} \right]    \right\}~,
\end{align}
and the decay rate for the on-shell photon corresponding to $q^2=0$  is given by ($F_{1}^{b}\equiv F_{1}^{b}(0)$)
\begin{align}
\label{eq:decay_rate_heavy}
\Gamma\left( h_{b} \to \eta_{b}\gamma \right) =  \frac{2Q_{b}^{2}}{3}\alpha_{\rm em}\frac{ m_{h_{b}}^{2} - m_{\eta_{b}}^{2}}{m_{h_{b}}}| F_{1}^{b} | ^{2}\,,
\end{align}
where, as before, we use $F_{1}^{b}\equiv F_{1}^{b}(0)$, for shortness.
We again neglect the disconnected contributions so that only the $b$-quark component of the electromagnetic current contributes to the matrix element and in Eq.~\eqref{eq:def_trans_ffb} one can replace
\begin{align}
J_{\rm em}^{\mu}(x) \mapsto J_{b}^{\mu}(x) = Q_{b}\,\bar{q}_{b}(x) \gamma^{\mu} q_{b}(x)\,.
\end{align}
In the rest frame of $h_{b}$ ($\mathbf{p}=0$) the three momentum $\mathbf{k}$ of the $\eta_{b}$ that corresponds to $q^{2}=0$ is 
\begin{align}
\label{eq:momentum_etab}
|\mathbf{k}| = \frac{ m_{h_{b}}^{2} -m_{\eta_{b}}^{2} }{2m_{h_{b}}} \simeq 488~\mathrm{MeV}\,, 
\end{align}
which differs from the momentum $\mathbf{k}$ of the $\eta_{c}$ meson in Eq.~(\ref{eq:momentum_etac}) by only $12~{\rm MeV}$.

Like we already mentioned in the Introduction, the $h_{b} \to \eta_{b}\gamma$ decay cannot be directly accessed from our lattices because the physical $b$-quark mass is too heavy and we therefore evaluate the $h_{b} \to \eta_{b}\gamma$ transition form factor $F_{1}^{b}$ by working with a series of heavy quark masses $m_{H}^{(n)}$ (with $am_{H}^{(n)} \lesssim 0.5$) and then extrapolate the results to the physical point corresponding to $m_{H} \to m_b$. 
In Tab.~\ref{tab:heavy_masses} we give the masses of simulated heavy quarks for each of the lattice ensembles used in this work. Note that the heavy quark masses $m_{H}^{(n)}$ satisfy 
\begin{align}
am_{H}^{(n+1)} = \lambda \, am_{H}^{(n)}  
\end{align}
with $\lambda=1.24283$, $n=1,\ldots , 6$, and $am_{H}^{(1)} = am_{c}$.~\footnote{At the heaviest simulated quark mass $m_{h}/m_{c} \simeq 3$ we decided to include the results on the C80 ensemble where $am_{H}^{(6)}\simeq 0.588$ is slightly larger than $0.5$.}
The parameter $\lambda$ is chosen in such a way that $m_{b}/m_{c} \simeq \lambda^{7} = 4.58$~\cite{FlavourLatticeAveragingGroupFLAG:2024oxs}.
\begin{table}[t]
\begin{ruledtabular}
\begin{tabular}{lcccccc}
\textrm{ID} & $am_{H}^{(1)}$ & $am_{H}^{(2)}$ & $am_{H}^{(3)}$ & $am_{H}^{(4)}$ & $am_{H}^{(5)}$ & $am_{H}^{(6)}$  \\
\colrule
\textrm{A48} & $0.26200$ & $0.32562$ & $0.40469$ &  $0.50296$ & // & // \\ 
\textrm{B64} & $0.23157$  & $0.28780$ & $0.35769$ & $0.44455$ & // & //  \\
\textrm{C80} & $0.19840$  & $0.24658$ & $0.30645$ &  $0.38087$ & $0.47336$ & $0.58830$ \\
\textrm{D96} & $0.16490$  &  $0.20494$ & $0.25471$ & $0.31656$ & $0.39343$ & $0.48897$  \\
\textrm{E112} & $0.14125$   & $0.17555$ & $0.21818$ & $0.27116$ & $0.33700$ & $0.41884$
\end{tabular}
\end{ruledtabular}
\caption{\small\sl Bare heavy quark masses $am_{H}^{(n)}$ used for the calculation of the $h_{b}\to\eta_{b}\gamma$ transition form factor, for all of the gauge ensembles used in this calculation, specified in Tab.~\ref{tab:simudetails}. \label{tab:heavy_masses}}
\end{table}
The lattice setup used to extract the transition form factor for each heavy quark mass $ m_{H}^{(n)} $ is essentially identical to that employed for $ h_{c} \to \eta_{c}\gamma $ and described in the previous Section. 

In the following we denote by $ \eta_{H} $ and $ h_{H} $ the fictitious states similar to $ \eta_{c} $ and $ h_{c} $ but with the heavy quark  $ m_{H} $ instead of $ m_{c} $. As before, for each heavy quark mass, the three-momentum $\mathbf{k}$ given to $ \eta_{H} $ is set to the value of Eq.~(\ref{eq:momentum_etab}) for each of the gauge ensembles, and not to that corresponding to the $h_{H}\to \eta_{H}\gamma$ decay. The difference between the transition form factors obtained by rescaling or not
rescaling the momentum $\mathbf{k}$ is entirely negligible, as we verified explicitly for $m_{H}=m_{c}$ where the deviation would be more pronounced (see Sec.~\ref{sec:num_res_hh}).

The interpolating operators of the $\eta_{H}$ ($\mathcal{O}_{\eta_{H}}$) and of the $h_{H}$ ($\mathcal{O}_{h_{H}})$ meson are defined as in Eqs.~(\ref{eq:interpolators}\,,\,\ref{eq:interpolators_TM}) with $q_{c}(x) \mapsto q_{H}(x)$. In order to improve the overlap with the ground state we tuned the number of smearing steps $n(m_{H})$ for each mass $m_{H}$ so as to ensure a smearing radius
\begin{align}
r_{0}(m_{H}) = \frac{an(m_{H})}{\sqrt{\kappa^{-1}+6}} \simeq \left( \frac{m_{c}}{m_{H}}\right) \times 0.15~\mathrm{fm}\,.
\end{align}
Like in the previous Section we compute the two point functions 
\begin{equation}
\label{eq:2pt_def_b}
\begin{split}
    C_{\text{2pt}}^{\eta_{H}}(t) &= \sum_{\mathbf{x}} e^{i\mathbf{k}\mathbf{x}} \langle 0 | \mathcal{O}_{\eta_{H}}(\mathbf{x}, t) \mathcal{O}_{\eta_{H}}^\dagger(\mathbf{0}, 0) | 0 \rangle, \\
    C_{\text{2pt}}^{h_{H}}(t) &= \sum_{\mathbf{x}} \langle 0 | \mathcal{O}_{h_{H}}(\mathbf{x}, t) \mathcal{O}_{h_{H}}^\dagger(\mathbf{0}, 0) | 0 \rangle~,
\end{split}
\end{equation}
in which the annihilation contributions have been neglected and extract 
$m_{h_{H}}$ and $E_{\eta_{H}}$, and the quantities
\begin{align}
\label{eq:heavy_MEM}
Z_{\eta_{H}} &= \langle 0 | \mathcal{O}_{\eta_{H}} | \eta_{H}(k) \rangle~,\qquad
Z_{h_{H}} = \langle h_{H}(p,\epsilon_{\lambda}) | \mathcal{O}^{\dagger}_{h_{H}} | 0 \rangle\,.
\end{align}
We also computed $C_{\text{2pt}}^{\eta_{H}}(t)$ with $\mathbf{k}=0$, which is needed to get $m_{\eta_{H}}$. Finally,  
the three point function corresponding to the connected diagram, needed for the determination of the transition form factor, is given by
\begin{widetext}
\begin{align}
\label{eq:three-point_H}
 C_{\text{3pt}}^{\mu}(t_{h}; t_{J}, m_{H}) &= \sum_{\mathbf{x}, \mathbf{y},\mathbf{z}} e^{i\mathbf{k}(\mathbf{x}-\mathbf{y})} \langle 0 | \mathcal{O}_{\eta_{H}}(\mathbf{x}, 0) J^{\mu}_{H}(\mathbf{y}, -t_{J}) \mathcal{O}_{h_H}^\dagger(\mathbf{z}, -t_{h}) | 0 \rangle_{\rm conn}~, \nonumber \\[8pt]
 &= 2\sum_{\mathbf{x},\mathbf{y},\mathbf{z}} \langle \tr \left[ S_{H}(x, z) \sigma^{12} S_{H}(z,y) \gamma^{\mu} S_{H}^{\theta}(y, x) \gamma^{5} \right]\rangle~,\end{align}
\end{widetext}
where $ J_{H}^{\mu}(x) = Q_{b}\,\bar{q}_{H}(x)\gamma^{\mu}q_{H}(x) $,  and $ S^{\theta}_{H}(x,y) $ is the heavy quark propagator computed on the gauge background $ U_{\mu}^{\theta}(x) $ as discussed in Eq.~\eqref{eq:twisted_gauge}. Note that  $ S_{H}(x,y) $ corresponds to $|\vec{\theta}|=0$. The twisting angle $\vec{\theta}$ is again chosen to produce the momentum specified in Eq.~(\ref{eq:momentum_etab}). 

\subsection{Numerical results for $m_{\eta_{H}}$ and $m_{h_{H}}$ \label{subsection:A}}
\begin{figure*}[t!]
\includegraphics[scale=0.40]{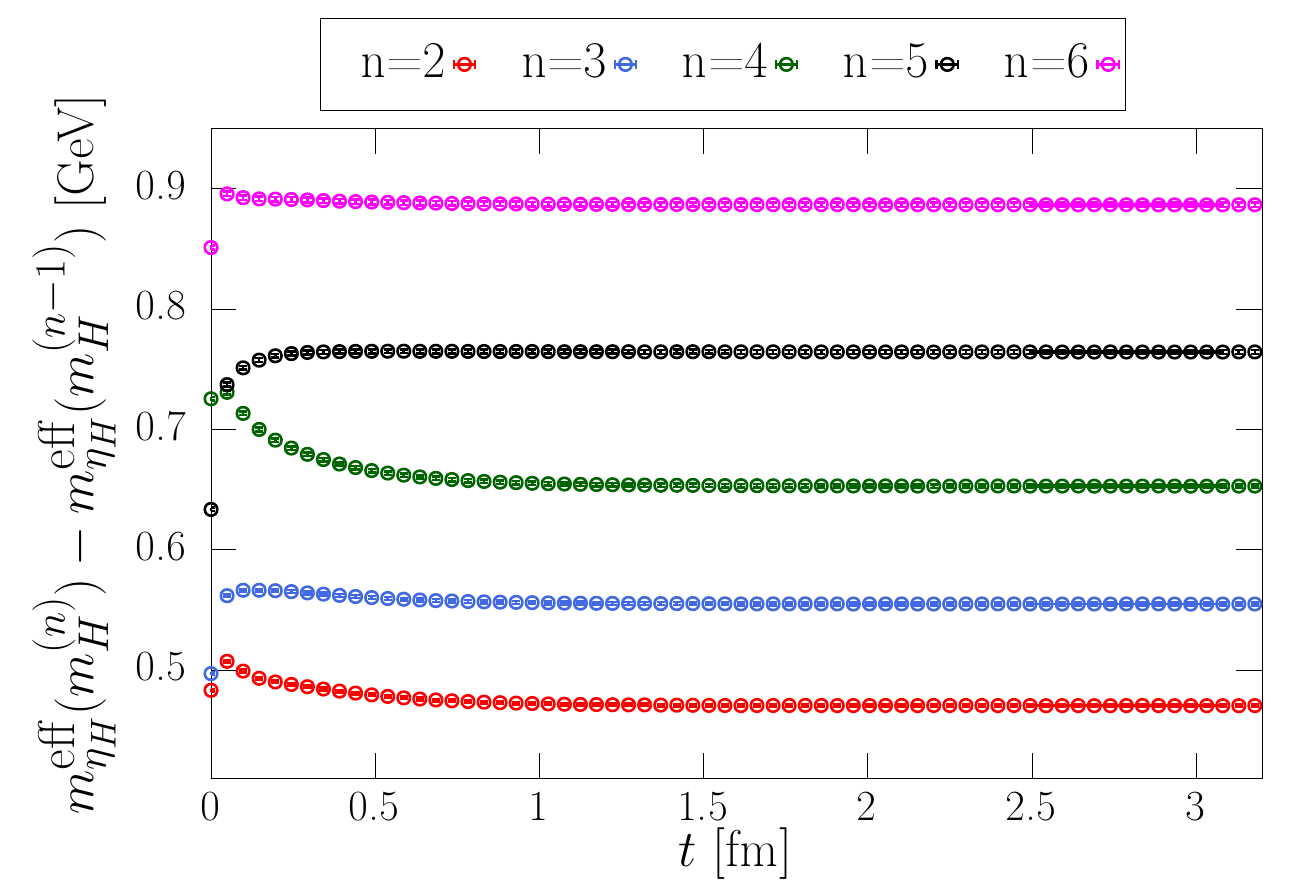}
\includegraphics[scale=0.40]{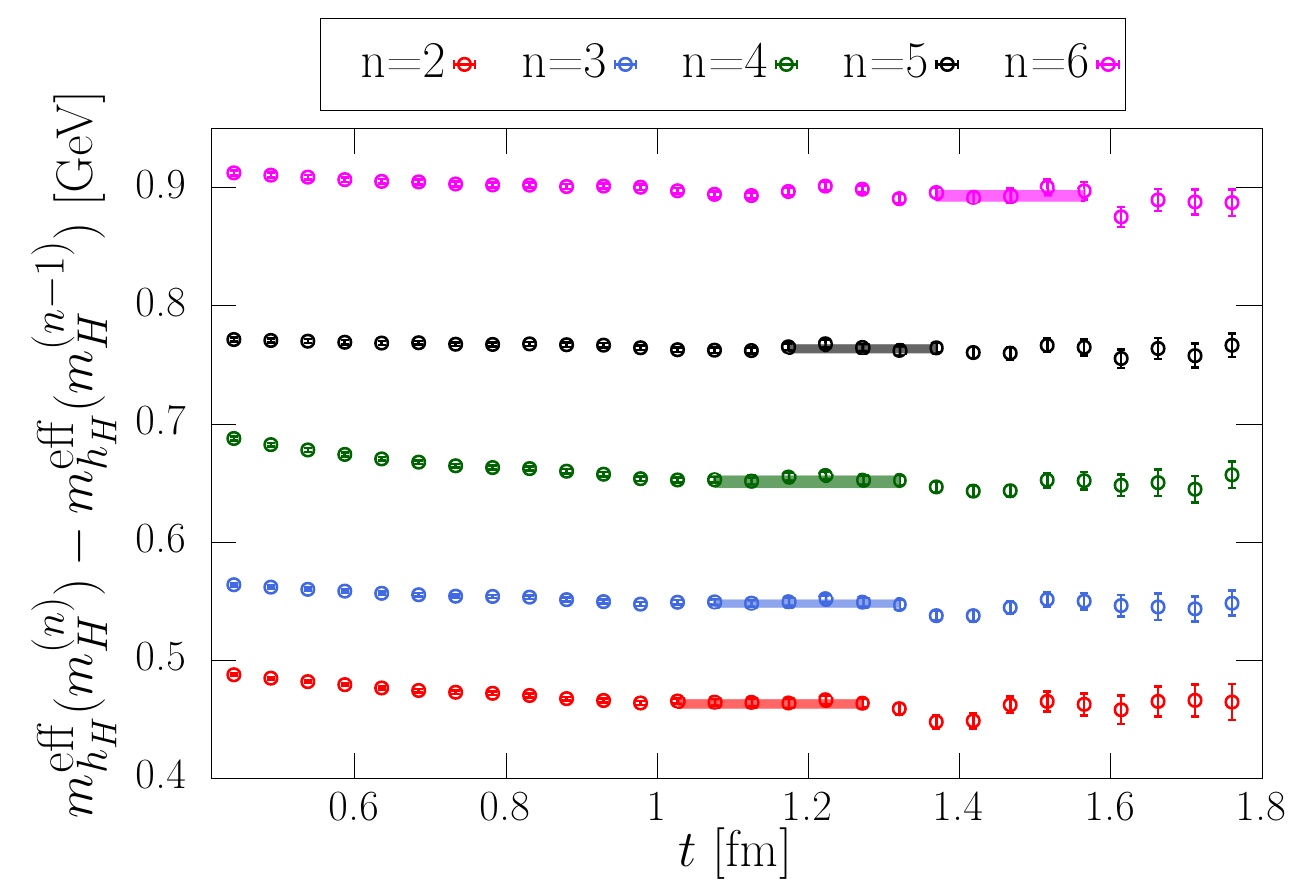}
\caption{\small\sl Difference between the effective mass of the $\eta_{H}$ (left) and $h_{H}$ meson (right), evaluated for two successive values of the heavy quark masses $m_{H}^{(n)}$ and $m_{H}^{(n-1)}$ for $n=2,3,4,5,6$. The illustrated results correspond to our finest lattice ensemble (E112 in Tab.~\ref{tab:simudetails}). \label{fig:etab_hb}}
\end{figure*}

For each of the heavy quark masses, listed in Tab.~\ref{tab:heavy_masses}, we compute the corresponding $m_{\eta_{H}}$ and $m_{h_{H}}$. The quality of the effective mass plateaus is illustrated in Fig.~\ref{fig:etab_hb} where we plot the difference between effective masses of the $\eta_{H}$  ($h_{H}$) at two consecutive heavy quark masses, $m_{H}^{(n)}$ and $m_{H}^{(n-1)}$. The results shown correspond to our determination on the finest lattice spacing ensemble employed in this calculation. From a fit to a constant for large time separations we  extract the mass differences,
\begin{equation}
\label{eq:eta_mass_diff}
\Delta m^{(n)}_{\eta_{H}} = m_{\eta_{H}}(m_{H}^{(n)}) - m_{\eta_{H}}(m_{H}^{(n-1)})\,,\qquad
\Delta m^{(n)}_{h_{H}} = m_{h_{H}}(m_{H}^{(n)}) - m_{h_{H}}(m_{H}^{(n-1)})\,,
\end{equation}
where we indicated with $m_{\eta_{H}}(m_{H})$ and $m_{h_{H}}(m_{H})$ the masses of the $\eta_{H}$ and $h_{H}$ meson evaluated at a given heavy-quark mass $m_{H}$.
As it can be appreciated in Fig.~\ref{fig:etab_hb}, the values of $\Delta m^{(n)}_{\eta_{H}}$ and $\Delta m^{(n)}_{h_{H}}$ are quite similar, consistent with what is known from experiment where $m_{h_{b}}^\mathrm{exp}-m_{h_c}^\mathrm{exp} \simeq 6.37$~GeV, is very close to $m_{\eta_{b}}^\mathrm{exp}-m_{\eta_c}^\mathrm{exp} \simeq 6.41$~GeV, thus differing by only $40$~MeV.

\begin{figure}
    \centering
    \includegraphics[width=0.95\linewidth]{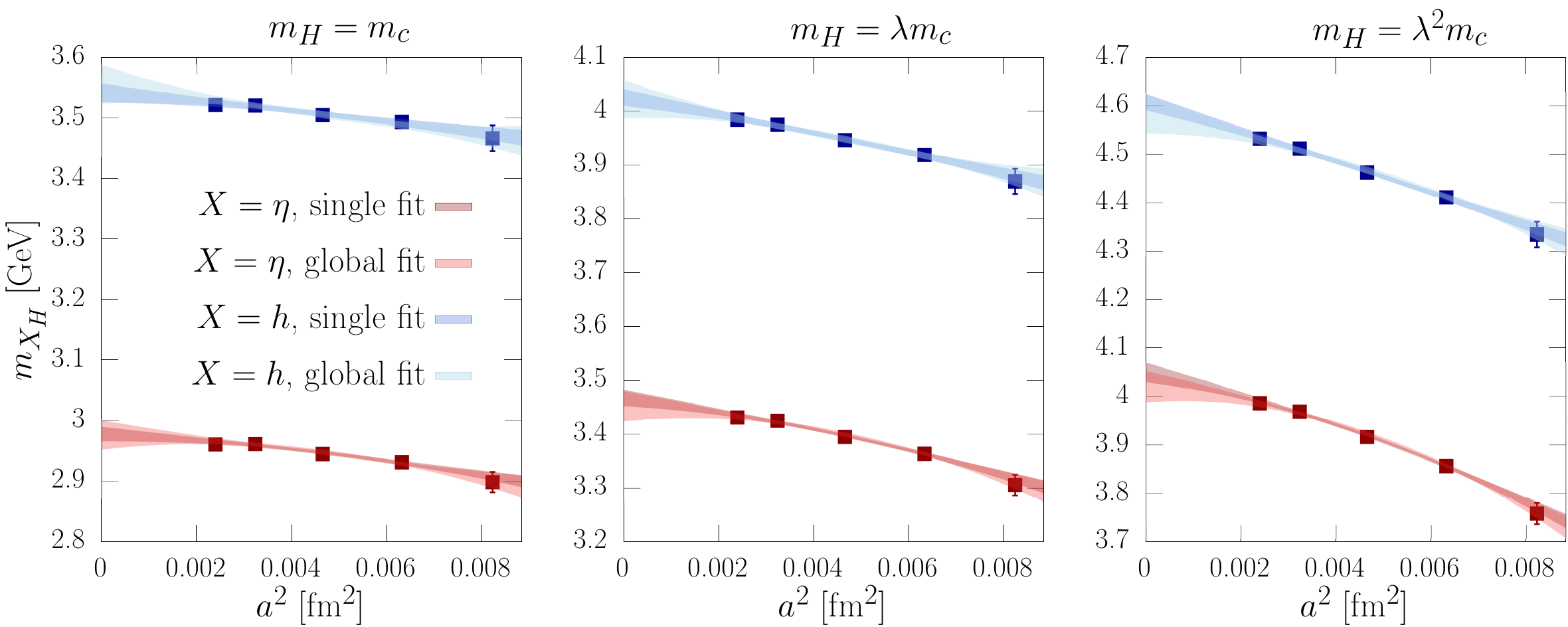}\\  \includegraphics[width=0.95\linewidth]{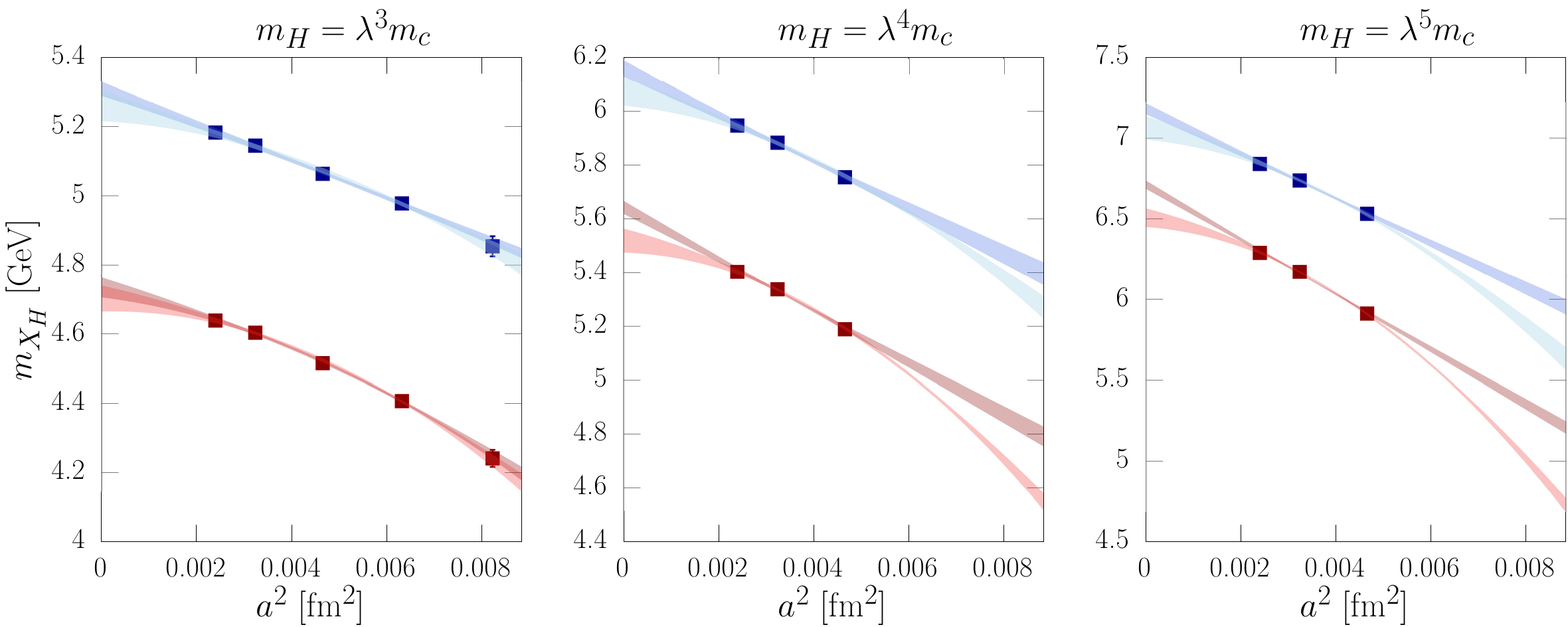}
    \caption{\small\sl Continuum extrapolation of the $\eta_{H}$ and $h_{H}$ masses, for all six simulated heavy quark masses $m_{H}$. In each plot, red and blue data points represent the lattice values of $m_{\eta_H}$ and $m_{h_H}$. The darker bands show the results of the separate $a^{2}$ and $a^{2}+a^{4}$ fits combined using the BAIC as described in the text. Note that for the two heaviest quark masses, where only results at three lattice spacings are available, we perform only the linear $a^{2}$ fit. The lighter bands correspond to the results of the global fit performed using Eq.~\eqref{eq:global_fit}. }
    \label{fig:masses_heavy}
\end{figure}

While the quality of the plateaus are always excellent for $\eta_{H}$, in the case of $h_{H}$ the S/N of $C_{\rm 2pt}^{h_{H}}(t)$ is expected to decrease over time as $e^{-(m_{h_{H}}-m_{\eta_{H}})t} \simeq e^{-(500~{\rm MeV}) t}$,  independently on the heavy quark mass $m_{H}$. However, when considering the correlated difference between the effective mass of $h_{H}$ computed with two consecutive heavy quark masses, the statistical errors partially cancel (because the S/N is largely independent on the heavy quark mass $m_{H}$ in the simulated region) and the quality of the plateaus substantially improves, allowing us to better identify the time interval where to perform a constant fit to our data. This is highlighted by the colored bands in the right panel of Fig.~\ref{fig:etab_hb}. The masses of the $h_{H}$ mesons are determined this way: we first evaluate on each ensemble the mass differences $\Delta m_{h_{H}}^{(n)}$ and then reconstruct $m_{h_{H}}(m_{H}^{(n)})$ as
\begin{align}
m_{h_{H}}(m_{H}^{(n)}) = m_{h_{c}} + \begin{cases}
    \displaystyle{\sum_{i=2}^{n} }\Delta m_{h_{H}}^{(i)}~, & \text{if} \qquad n \geq 2~\\
    \quad\qquad 0~,              & \text{otherwise}
\end{cases}\,.
\end{align}

To extrapolate $m_{\eta_{H}}$ and $m_{h_{H}}$ to the continuum limit for each simulated heavy quark mass $m_{H}$, we follow two different strategies. In the first, we take the continuum limit of both $m_{\eta_{H}}$ and $m_{h_{H}}$, for each value of the heavy-quark mass $m_{H}$, by using the ansatz
\begin{equation}
\begin{split}
m_{\eta_{H}}(m_{H}, a) &= A_{0}(m_{H}) + A_{1}(m_{H})\,a^{2} + A_{2}(m_{H})\,a^{4}\,, \\[8pt]  
m_{h_{H}}(m_{H},a) &= B_{0}(m_{H}) + B_{1}(m_{H})\,a^{2} + B_{2}(m_{H})\,a^{4}\,,
\end{split}
\end{equation}
where distinct fit parameters, $A_{0}(m_{H})$, $A_{1}(m_{H})$, $A_{2}(m_{H})$, $B_{0}(m_{H})$, $B_{1}(m_{H})$, $B_{2}(m_{H})$ are used for each $m_{H}$ value. We perform both linear and quadratic fits in $a^{2}$, which are then combined using the BAIC. For the two heaviest heavy quark mass $m_{H} = \lambda^{4}m_{c}$ and $m_{H} = \lambda^{5}m_{c}$, where the results only at three lattice spacings are available (C80, D96 and E112) we perform a simple linear $a^{2}$ fit ($A_{2}(m_{H})=B_{2}(m_{H})=0$). We never impose a prior on any of the fit parameters.

Our second strategy is to perform a global fit that simultaneously describes the dependence on lattice artifacts according to:
\begin{equation}
\label{eq:global_fit}
\begin{split}
m_{\eta_H}(m_H, a) 
  &= A'_0(m_H) \times \Bigl( 1 
      + A'_1\,a^2 
      + A'_{1,m}\,a^2\,m_H^2 
      + A'_2\,a^4 
      + A'_{2,m}\,a^4\,m_H^2 
      + A'_{2,m_2}\,a^4\,m_H^4 
  \Bigr),\\[8pt]
m_{h_H}(m_H, a) 
  &= B'_0(m_H) \times \Bigl( 1 
      + B'_1\,a^2 
      + B'_{1,m}\,a^2\,m_H^2 
      + B'_2\,a^4 
      + B'_{2,m}\,a^4\,m_H^2 
      + B'_{2,m_2}\,a^4\,m_H^4 
  \Bigr),
\end{split}
\end{equation}
where $A'_0(m_H)$ and $B'_0(m_H)$ are fit parameters specific for each heavy quark mass, while the remaining parameters $A'_{1,2}$ and $B'_{1,2}$ describe the cutoff effects and are common to all the simulated heavy quark masses. This form already provides a good description of the data with $A'_{2,m_{2}} = B'_{2,m_{2}} = 0$.

Illustration of how these two strategies compare is provided in Fig.~\ref{fig:masses_heavy}. In each plot, red and blue data points represent the lattice values of $m_{\eta_H}$ and $m_{h_H}$, respectively. The darker bands show the separate continuum extrapolations. The quality of the fit is excellent, with at least one of the linear or quadratic fits in $a^2$ giving a reduced $\chi^2$ close to unity. The lighter bands represent the corresponding results of the global fits (with $A'_{2,m_{2}} = B'_{2,m_{2}} = 0$). When five lattice spacings are available for a given $m_H$, the two strategies agree very well. However, when only three lattice spacings are available (for the two heaviest quark masses), the results for $m_{\eta_{H}}$, obtained using the two strategies, differ by a few percent (we also note that the $\chi^{2}/\mathrm{dof}$ of the linear fit, for the heaviest simulated heavy quark mass, is not very good, about $2.5$). 

For the first four simulated heavy quark masses we take as our final results for $m_{\eta_{H}}$ and $m_{h_{H}}$ the determinations obtained performing separate fits for each heavy-quark mass. For the two heaviest masses, $m_H = \lambda^4 m_c$ and $m_H = \lambda^5 m_c$, while we consider more solid the results of the global fit, we decided to stay on the more conservative side and combine the results of both strategies using Eq.~(\ref{eq:BMA_ave}), assigning equal weights to the two results in the continuum limit. In Table~\ref{tab:heavy_hadron_masses}, we give $m_{\eta_H}$ and $m_{h_H}$ in the continuum limit for each simulated heavy quark mass $m_H$.

\begin{table}[]
\begin{ruledtabular}
\begin{tabular}{lcccccc}
 & $m_{H}^{(1)}$ & $m_{H}^{(2)}$ & $m_{H}^{(3)}$ & $m_{H}^{(4)}$ & $m_{H}^{(5)}$ & $m_{H}^{(6)}$  \\
\colrule
$m_{\eta_{H}}~[\rm{GeV}]$ & $2.978(12)$ & $3.467(15)$ & $4.049(20)$ &  $4.735(29)$ & $5.581(70)$ & $6.61(11)$ \\[6pt]
$m_{h_{H}}~[\rm{GeV}]$ & $3.541(15)$  & $4.025(16)$ & $4.607(17)$ & $5.309(20)$ & $6.116(58)$ & $7.123(77)$  
\end{tabular}
\end{ruledtabular}
\caption{\small\sl Masses of $\eta_{H}$ and $h_{H}$ after extrapolation to the continuum limit. Note that the heavy quark mass $m_{H}^{(n)} = \lambda^{n} m_{c}$, with $\lambda=1.24283$. \label{tab:heavy_hadron_masses}}
\end{table}

\subsection{Numerical results for $h_{H} \to \eta_{H}\gamma$}
\label{sec:num_res_hh}
We now discuss the determination of the amplitude for the process $h_{H} \to \eta_{H}\gamma$. 
Like in the previous Section we compute the three point correlation function $C_{\rm 3pt}^{\mu}(t_{h}; t_{J}, m_{H})$ for every 
heavy quark mass $m_{H}^{(n)}$ given in Table~\ref{tab:heavy_masses} and on all of the ETMC ensembles of gauge field configurations, cf. Table~\ref{tab:simudetails}. For each $m_{H}^{(n)}$ we fix the three-momentum $\mathbf{k}$ of the $\eta_{H}$ to 
the value indicated in Eq.~(\ref{eq:momentum_etab}),  corresponding to the physical momentum of the $\eta_{b}$ in the
$h_{b}\to \eta_{b}\gamma$ decay. This choice does not precisely match that of the 
$h_{H}\to \eta_{H}\gamma$ decay, so it corresponds to a non-zero photon off-shellness $q^{2}$ for $m_{H} < m_{b}$.  This, however, is not a source of problems because at the physical point $m_{H} = m_{b}$, the off-shellness $q^{2}$ extrapolates to its correct, vanishing, value. 
Furthermore, by comparing the results for $m_{H}=m_{c}$ obtained using the three-momentum $\mathbf{k}$ corresponding to 
$h_{c}\to \eta_{c}\gamma$ with that corresponding to  $h_{b}\to \eta_{b}\gamma$, the 
difference between rescaling or not $\bs{k}$ is far below our uncertainties 
and therefore completely negligible. In what follows, we denote by $F_{1}(m_{H})$ the form 
factor extracted from the three-point function 
$C_{\rm 3pt}^{\mu}(t_{H}; t_{J}, m_{H})$ with $\mathbf{k}$ set as in 
Eq.~(\ref{eq:momentum_etab}).

On each gauge ensemble, instead of directly determining the form factor $F_{1}(m_{H})$ for each simulated value of $m_{H}$, we find it useful to consider the ratios of form factors evaluated at two successive values, $m_{H}^{(n)}$ and $m_{H}^{(n-1)}$, of the heavy-quark mass, namely 
\begin{align}
\label{eq:ratio}
r_{\lambda}(m_{H}^{(n)})  \equiv F_{1}(m_{H}^{(n)})/F_{1}(\lambda^{-1} m_{H}^{(n)})\,. 
\end{align}
The ratios $r_{\lambda}(m_{H})$ can be extracted from the following estimator built in term of the ratio of three-point functions
\begin{align}
\label{eq:ratio_estimator}
\overline{r}^{(n)}_{\lambda}(t_{h}; t_{J}) &= R_{h\eta }^{(n)} ~ \exp\left\{\Delta E_{\eta_{H}}^{(n)} \, t_{J}\right\} \exp\left\{\Delta m_{h_{H}}^{(n)}\, (t_{h}-t_{J})\right\} \frac{C_{\rm 3pt}^{1}(t_{h};t_{J}, m_{H}^{(n)} )}{ C_{\rm 3pt}^{1}(t_{h};t_{J}, m_{H}^{(n-1)}) }~, \\
\label{eq:limit_ratio}
r_{\lambda}(m_{H}^{(n)}) &= \lim_{ t_{J} \to \infty} \,\lim_{t_{h} - t_{J} \to \infty} \overline{r}_{\lambda}^{(n)}(t_{h}; t_{J})\,,
\end{align}
where
\begin{align}
\Delta E^{(n)}_{\eta_{H}} = E_{\eta_{H}}(m_{H}^{(n)}) - E_{\eta_{H}}(m_{H}^{(n-1)}) ~, \qquad
R_{h\eta}^{(n)} = \frac{E_{\eta_{H}}(m_{H}^{(n)})}{ E_{\eta_{H}}(m_{H}^{(n-1)})
} \times \frac{Z_{\eta_{H}}(m_{H}^{(n-1)}) \, }{Z_{\eta_{H}}(m_{H}^{(n)})} \times \frac{Z_{h_{H}}(m_{H}^{(n-1)})}{ Z_{h_{H}}(m_{H}^{(n)})}~,
\end{align}
already defined in Eq.~(\ref{eq:heavy_MEM}) at a given $m_{H}$. 

Considering ratios of form factors at nearby masses provides several advantages. First, in the ratios statistical errors are reduced and only the mass difference $\Delta m_{h_{H}}^{(n)}$ enters the exponential terms in Eq.~(\ref{eq:ratio_estimator}). As discussed above, the uncertainties of the differences of effective $h_{H}$-masses are reduced and the plateaus are more stable compared to that of the $h_{H}$ mass itself. Furthermore, the mass independent cutoff effects cancel in the ratio, thereby reducing the lattice artifacts. We therefore compute the ratios $r_{\lambda}(m_{H}^{(n)})$ on each gauge ensemble, and extrapolate them to the continuum limit. Then, having extracted $F_{1}(m_{c})$ by the same procedure 
described in Sec.~\ref{sec:hc_etac} (but fixing the $\eta_{c}$ three-momentum to 
Eq.~\eqref{eq:momentum_etab} rather than Eq.~\eqref{eq:momentum_etac}), we reconstruct 
$F_{1}(m_{H}^{(n)})$ as
\begin{align}
\label{eq:ff_reco}
F_{1}(m_{H}^{(n)}) 
= F_{1}(m_{c}) \times \prod_{i=2}^{n} r_{\lambda}(m_{H}^{(i)})~, 
\qquad n > 1\,.
\end{align}

The idea of considering ratios of form factors at consecutive heavy quark masses is not new. It was introduced in Ref.~\cite{ETM:2009sed} in the context of heavy-light systems, and it is nowadays known as the \textit{ratio-method}. It has been successfully applied to the calculation of a variety of $B$-physics observables, including the b-quark mass $m_{b}$~\cite{ETM:2011zey,ETM:2016nbo},  the $B^{(*)}$- and $B_{s}^{(*)}-$meson decay constant~\cite{ETM:2011zey, ETM:2013jap, ETM:2016nbo, Balasubramamian:2019wgx}, $B_s\to D_s$ semileptonic form factors~\cite{Atoui:2013zza,Blossier:2021xvl}, and the bag parameter describing $B^{0}-\bar{B}^{0}$ mixing~\cite{ETM:2013jap}. For the heavy-light systems the ratio-method allows one to turn the extrapolation $m_{H}\to m_{b}$, to an interpolation. Indeed, owing to the fact that heavy quark effective theory (HQET) predicts asymptotic power law scaling for the observables related to heavy-light hadrons $O_{lh}$, namely $O_{lh} \propto m_{H}^{\alpha}$ up to corrections proportional to powers of $\Lambda/m_{H}$, 
by constructing the following ratio:
\begin{align}
r^{O_{hl}}_{\lambda}(m_{H}) = \lambda^{-\alpha} \frac{O_{hl}(m_{H})}{{O}_{hl}(m_{H}/\lambda)} ~,
\end{align}
one has that $\lim_{m_{H}\to \infty} r^{\mathcal{O}_{hl}}_{\lambda}(m_{H}) =1$, thereby  turning the extrapolation to the physical point $m_{H}\to m_{b}$ into an interpolation ($r^{O_{hl}}_{\lambda}(\infty) = 1$). Having parametrized the mass dependence of $r_{\lambda}^{O_{hl}}(m_{H})$ one can then obtain $O_{hl}(m_{b})$ from the knowledge of $O_{hl}(m_{c})$ provided that $\lambda^{N} = m_{b}/m_{c}$ for some integer $N$. 

However, in the case of heavy quarkonia, the presence of different energy scales in (potential) NRQCD, and the fact that both charmonium and bottomonium systems may exhibit very distinct behavior relative to the one expected in the asymptotic limit, $m_{H}\to \infty$, implies that 
a simple scaling law valid from $m_{H}\simeq m_{c}, m_{b}$ up to $m_{H} \to \infty$ might not hold.
For that reason, in this work we employ the ratio method only as a tool to reduce the statistical uncertainties and cutoff effects. After continuum extrapolation, we will determine the form factors employing Eq.~(\ref{eq:ff_reco}) and discuss in the next Section the extrapolation of $F_{1}(m_{H})$ to the physical point $m_{H}\to m_{b}$.

\begin{figure}[]
    \centering
    \includegraphics[width=0.95\linewidth]{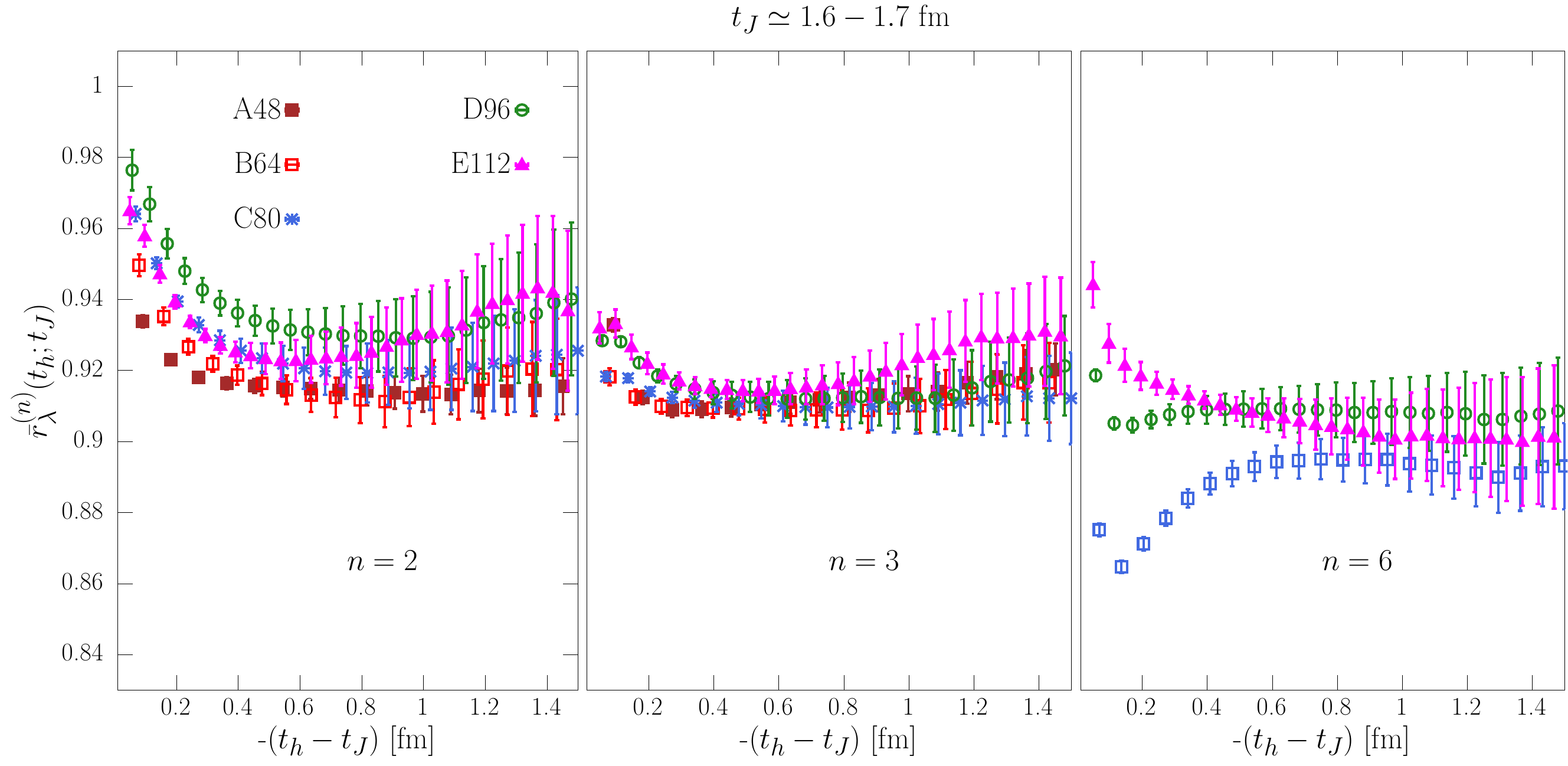}
    \caption{\small\sl The estimator $\bar{r}_{\lambda}^{(n)}(t_{h};t_{J})$ of the ratio $r_{\lambda}(m_{H}^{(n)})$ for $n=2$ (left), $n=3$ (middle) and $n=6$ (right), for all the gauge ensembles used for the present computation, and for fixed $t_{J}\simeq 1.6-1.7~\mathrm{fm}$. For $n=2,3$ results at five lattice spacings are available while for $n=6$ the ratio is computed on three lattice spacings (C80, D96, E112).}
    \label{fig:ratio}
\end{figure}

In order to reduce the UV cutoff effects induced by the local vector current of the heavy quark ($J_{H}^{\mu}(x)$), we find it useful to modify the estimator $\bar{r}_{\lambda}^{(n)}(t_h; t_{J})$ as
\begin{align}
\bar{r}_{\lambda}^{(n)}(t_h; t_{J}) \to \Delta_{\rm RC}^{(n)} \, \bar{r}_{\lambda}^{(n)}(t_h; t_{J})~,
\end{align}
where 
\begin{align}
\Delta_{\rm RC}^{(n)} &= \frac{ Z^{\rm WI}_{V}(m_{H}^{(n)})}{ Z^{\rm WI}_{V}(m_{H}^{(n-1)})}= 1 + O(a^{2}m_{H}^{2})~,\
\end{align}
and where $Z_{V}^{\rm WI}(m_{H})$ is the renormalization constant of the local vector current of the heavy quark,
determined from the twisted-mass Ward identity as
\begin{align}
\label{eq:WI_TM}
Z_{V}^{\rm WI}(m_{H}) = \lim_{t\to\infty}2m_{H}\,\frac{ \displaystyle{\sum_{\mathbf{x}}} \langle \mathcal{O}_{\eta_{H}}(\mathbf{x},t)\, \bar{q}_{H}(0) \gamma^{5} q'_{H}(0) \rangle}{\partial_{t}\displaystyle{\sum_{\mathbf{x}}} \langle \mathcal{O}_{\eta_{H}}(\mathbf{x},t)\, \bar{q}_{H}(0) \gamma^{5} \gamma^{0} q'_{H}(0)  \rangle }~.
\end{align}
In Eq.~(\ref{eq:WI_TM}), the primed field $q'_{H}(x)$ differs from the field $q_{H}(x)$ by the sign of the twisted Wilson term (see the discussion around Eq.~(\ref{eq:interpolators_TM})). In Fig.~\ref{fig:ratio}, we show the estimators $\bar{r}_{\lambda}^{(n)}(t_{h};t_{J})$ for fixed $t_{J} \simeq 1.6-1.7~{\rm fm}$, for $n=2,3,6$ and for all gauge ensembles employed for the calculation.
\begin{figure}[!]
    \centering    \includegraphics[width=0.93\linewidth]{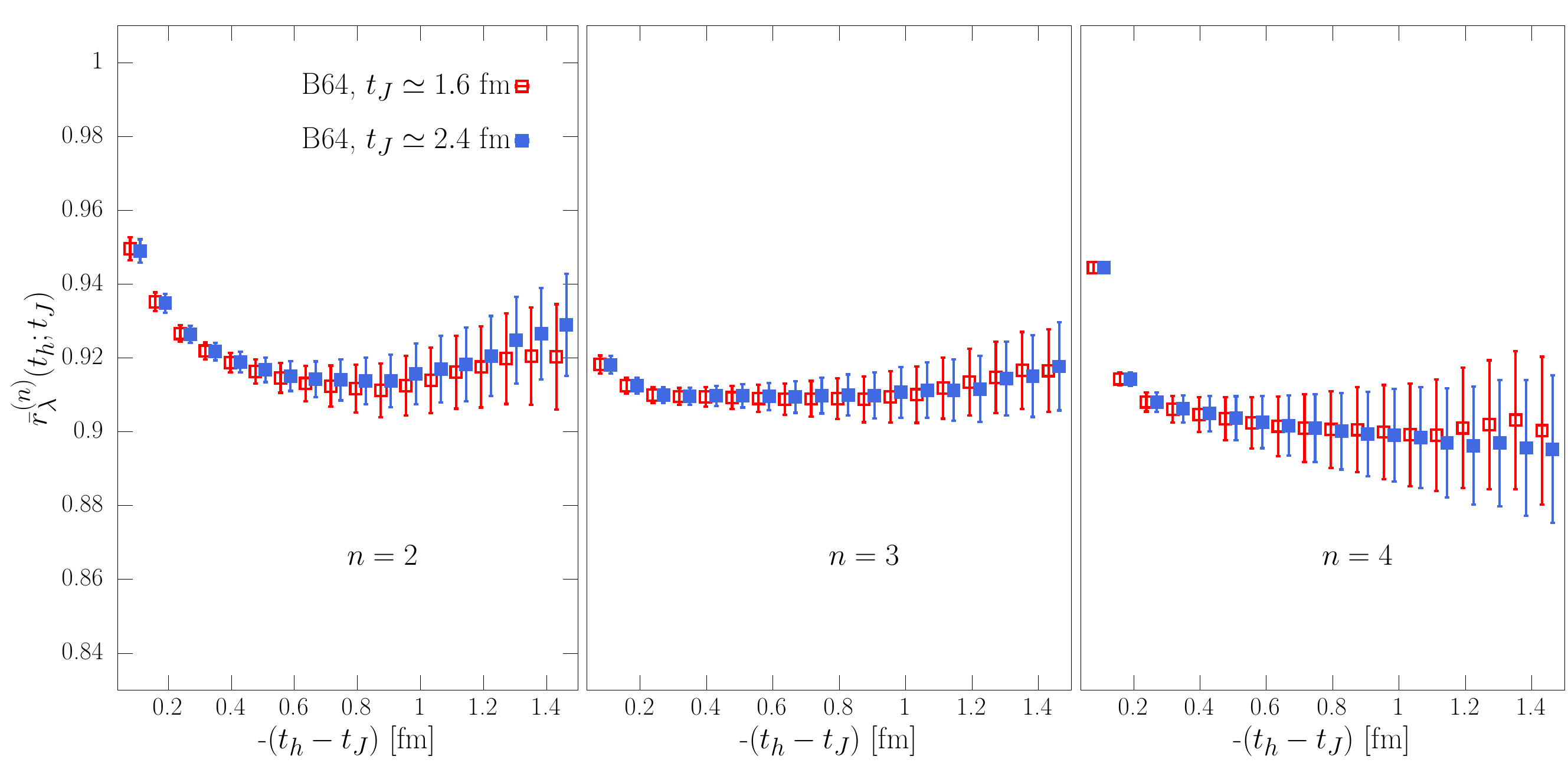}
    \caption{\small\sl Comparison between the results obtained for the the estimators $r_{\lambda}^{(n)}(t_{h}; t_{J})$ with $n=2,3,4$ on the B64 ensemble for two different values of $t_{J}$, $t_{J}\simeq 1.6~\mathrm{ fm}$ and $t_{J}\simeq 2.4~\mathrm{ fm}$.}
    \label{fig:comp_tj_B64}
\end{figure}
As the figure shows, like in the case of $F_{1}^{c}$, we find the cutoff effects to be remarkably small.  The ratios $r_{\lambda}(m_{H}^{(n)})$ are always smaller than one, about $0.9$, and only smoothly dependent on $n$. Therefore, with increasing heavy quark mass the form factor $F_{1}(m_{H})$ is decreasing, roughly by about $10\%$ every after each increment of $m_{H}$ by a factor of $\lambda = 1.24283$. Similarly to the case of $h_{c}\to \eta_{c}\gamma$, we carried out a study of the $t_{J}-$dependence of our results by computing the three-point functions $C_{\rm 3pt}^{\mu}(t_{h};t_{J}, m_{H})$ on the B64 ensemble for a second value of $t_{J}\simeq 2.4~\mathrm{fm}$. 
In Fig.~\ref{fig:comp_tj_B64} we show the comparison between the results obtained with $t_{J}\simeq 1.6~\mathrm{ fm}$ and those obtained with $t_{J}\simeq 2.4~\mathrm{ fm}$ for the ratios $r_{\lambda}(m_{H}^{(n)})$, with $n=2,3,4$. As the figure shows, no differences are visible and, as in the case of $h_{c}\to\eta_{c}\gamma$, the statistical noise does not deteriorate while going for larger $t_{J}$, in line with expectations discussed in Sec.~\ref{sec:extr_form_factor_charm}. 
On each gauge ensemble we extract the ratios $r_{\lambda}(m_{H}^{(n)})$ by fitting $\bar{r}_{\lambda}^{(n)}(t_{h};t_{J})$ to a constant in the region where a plateau is visible. 
\begin{figure}[!]
    \centering
    \includegraphics[width=0.93\linewidth]{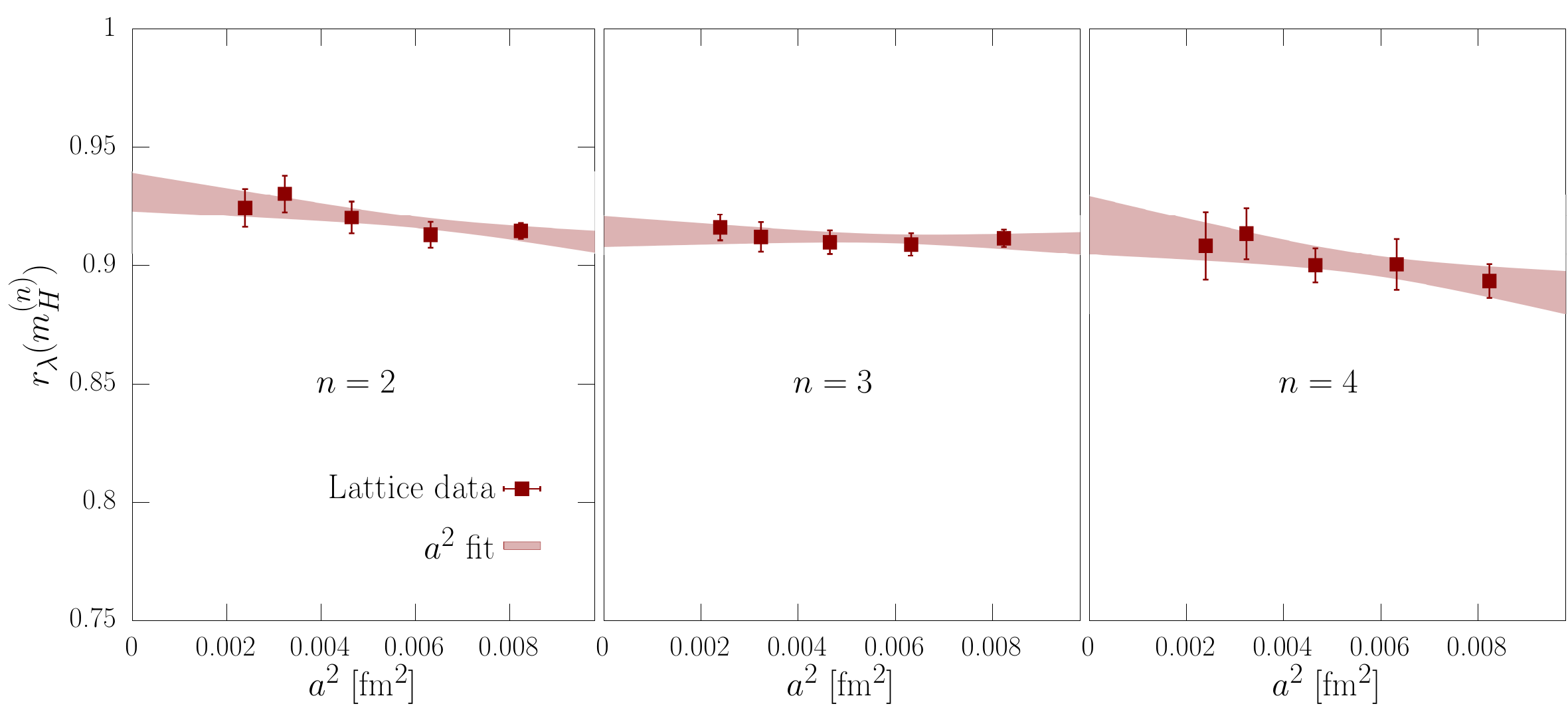} \\
    \includegraphics[width=0.93\linewidth]{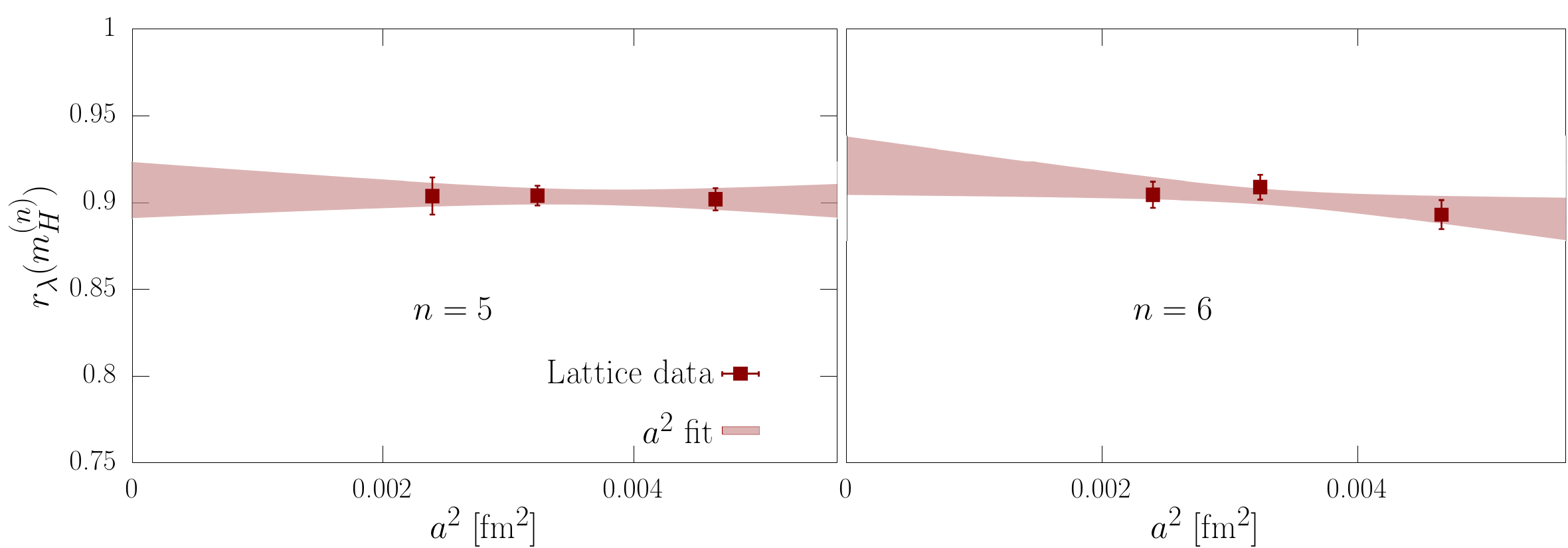}
    \caption{\small\sl Continuum-limit extrapolation for the  ratios $r_{\lambda}(m_{H}^{(n)})$ with $n=2$ (top-left), $n=3$ (top-center), $n=4$ (top-right), $n=5$ (bottom-left) and $n=6$ (bottom-right). The extrapolation is performed through a linear $a^{2}$ fit to the lattice data. In all cases we have $\chi^{2}/\mathrm{dof} \lesssim 1$.}
    \label{fig:continuum_2_4}
\end{figure}
Each of $r_{\lambda}(m_{H}^{(n)})$ now should be extrapolated to the continuum limit. Given the smallness of the UV cutoff effects in the ratios, we extrapolate our lattice data to the continuum limit through a simple linear fit in $a^{2}$, which is shown in Fig.~\ref{fig:continuum_2_4}. 
Contrary to the masses of $\eta_{H}$ and $h_{H}$, where $a^{4}$ cutoff effects are clearly visible, in this case the lattice spacing dependence is very mild. The reduced $\chi^{2}$ of the linear fits is always close to (or smaller than) unit. We also checked that the ratio-method applied to the masses ($m_{\eta_H}$ and $m_{h_H}$) gives results in the continuum limit fully consistent with those we presented in~\ref{subsection:A}.
\begin{figure}[!]
    \centering
    \includegraphics[width=0.7\linewidth]{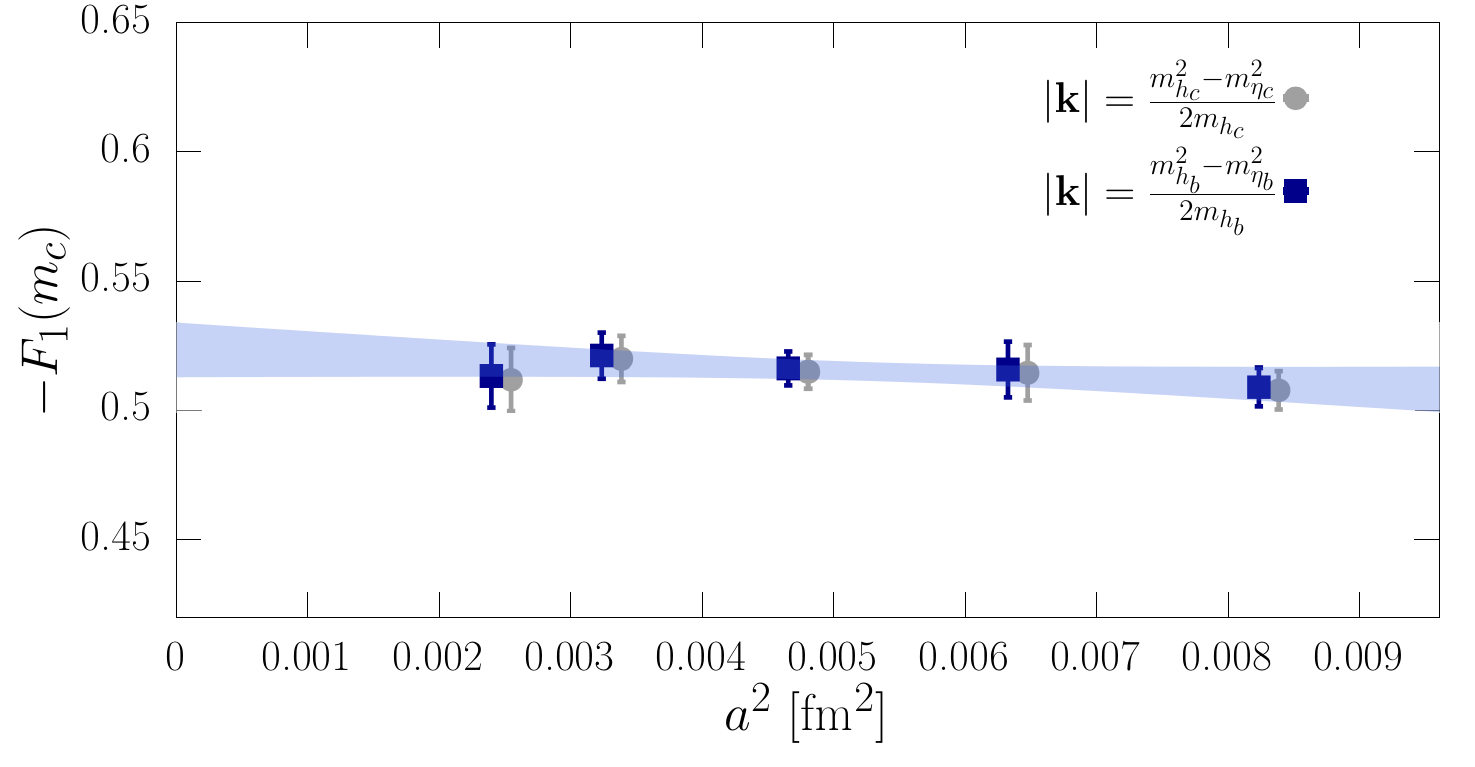}
    \caption{\small\sl Linear extrapolation of $F_{1}(m_{c})$ in $a^{2}$ to the continuum limit (colored band). Data points obtained by fixing $|\bf{k}|$ by the physical masses of bottomia (squares) are compared to those obtained in the previous section in which $|\bf{k}|$ was fixed by the physical masses of charmonia (circles) and slightly offset for easier comparison.}
    \label{fig:comp_etac}
\end{figure}
The last step of this analysis is the determination of the form factor $F_{1}(m_{c})$. Whether we use $|\bf{k}|$ from the physical charmonia or bottomia, the resulting $F_{1}(m_{c})$ remains practically indistinguishable, as displayed in Fig.~\ref{fig:comp_etac}.~\footnote{Obviously, $F_{1}(m_{c})$ with $|\bf{k}|$ fixed by using the physical masses of charmonia coincides with $F_1^c$ discussed in the previous Section. }
Finally, we obtain $F_{1}(m_{H})$ by multiplying $F_{1}(m_{c})$ and the ratios $r_{\lambda}(m_{H}^{(n)})$, cf. Eq.~\eqref{eq:ff_reco}. The results are reported in Tab.~\ref{tab:FF}. We are now going to describe how to extrapolate the form factor to the physical $b$-quark mass. 
\begin{table}[t]
\begin{ruledtabular}
\begin{tabular}{lcccccc}
 & $m_{H}^{(1)}$ & $m_{H}^{(2)}$ & $m_{H}^{(3)}$ & $m_{H}^{(4)}$ & $m_{H}^{(5)}$ & $m_{H}^{(6)}$  \\
\colrule
$-F_{1}(m_{H}^{(n)})$  & $0.523(10)$ & $0.4872(91)$ & $0.4455(83)$ &  $0.4086(89)$ & $0.371(10)$ & $0.342(13)$ \\[6pt]
$\,\,\,\,\,r_{\lambda}(m_{H}^{(n)})$ &  & $0.9311(81)$ & $0.9144(64)$ & $0.917(12)$ & $0.907(16)$ & $0.921(17)$  
\end{tabular}
\end{ruledtabular}
\caption{\small\sl Values of the form factor $F_{1}(m_{H}^{(n)})$ and of the ratio $r_{\lambda}(m_{H}^{(n)})$ for each simulated value of the heavy-quark mass $m_{H}^{(n)} = \lambda^{n-1} m_{c}$, with $\lambda=1.24283$. \label{tab:FF}}
\end{table}

\subsection{Extrapolation $m_{H}\to m_{b}$}
\label{sec:subsec_NRQCD}
Electric-dipole induced transitions (E1) of heavy quarkonia, such as $h_{b} \to \eta_{b}\gamma$, have been studied in Ref.~\cite{Brambilla:2012be} using NRQCD~\cite{Caswell:1985ui,Bodwin:1994jh}, and in potential NRQCD (pNRQCD)~\cite{Pineda:1997bj,Brambilla:1999xf,Segovia:2018qzb}. 
In Ref.~\cite{Brambilla:2012be} it is found that the leading order expression for the decay 
rate $\Gamma(h_{H}\to \eta_{H}\gamma)$, as a function of the quark mass $m_{H}$ reads~\footnote{In NRQCD the heavy quark mass, appearing in Eq.~\eqref{eq:scaling_width},  
should be identified with the quark pole mass. }
\begin{align}
\label{eq:scaling_width}
\Gamma(h_{H}\to \eta_{H}\gamma) \propto \frac{|\mathbf{k}|^{3}}{(m_{H})^{2} v_{H}^{2}} \left[ 1 + O(
m^{-1}_{H}, v_{H}^{2})  \right]~,
\end{align}

where the heavy quark velocity, $v_{H}\ll 1$, is a dynamical quantity, function of the strong interaction scale $\Lambda_{\rm QCD}$. In contrast to the heavy-light systems, in NRQCD the presence of different scales, $\Lambda/m_{H}$, $v_{H}$, $\ldots$, and of their interplay, does not allow one to obtain a simple scaling relation. 

For heavy quarkonia, two regimes are typically distinguished: the weakly coupled regime $m_{H}v_{H}^{2} \gg \Lambda$, and the strongly coupled one $m_{H} v_{H}^{2} \ll \Lambda$.
The weakly coupled regime is approached in the asymptotic limit $m_{H}\to \infty$, in which the quark-antiquark system is mainly sensitive to the short distance Coulombic part of the quark-antiquark potential $V_{H}(r)$, given (up to higher-order corrections in $\alpha_{s}$) by
\begin{align}
V_{H}^{\rm pert}(r) = -\frac{C_{F}\alpha_{s}(r)}{r}\,,
\end{align}
with $C_{F}=4/3$ for QCD. In that situation, the following scaling law applies
\begin{align}\label{eq:scaling0}
v_{H}(m_{H}) \propto \alpha_{s}(m_{H}v_{H})~, \qquad 
|\bs{k}| \simeq \Delta m_{h\eta} &\propto m_{H}v_{H}^{2} \propto  m_{H}\alpha_{s}^{2}(m_{H}v_{H})~,
\end{align}
where $\Delta m_{h\eta} \equiv m_{h_{H}} - m_{\eta_{H}}$.
From the comparison between Eq.~(\ref{eq:scaling_width}) and Eq.~(\ref{eq:decay_rate_heavy}), it follows that in the Coulombic limit the form factor $F_{1}(m_{H})$ scales as $F_{1}(m_{H}) \propto \alpha_{s}(m_{H}\, v_{H})$. 
Instead, in the strongly coupled regime the system is also sensitive to the long distance, nonperturbative, part of the quark-antiquark potential $V_{H}(r) \simeq \sigma r$.

Determining whether a given quarkonium state is weakly or strongly coupled is far from obvious. The ground states, such as $\eta_{c,b}$, are more likely to be weakly coupled: their wave functions are relatively localized around the system's center of mass and therefore they are likely to be more sensitive to the short distance part of the potential $V_{H}(r)$. In contrast, the excited states, such as $h_{c,b}$, tend to be more strongly coupled: their wave functions are more spread out and thus increasingly sensitive to the nonperturbative part of the potential.

In our case, we already noted that the photon momentum $-\mathbf{k}$ only slightly changes when going from $m_{c}$ to $m_{b}$ (by about $2\%$). Thus, considering $\mathbf{k}$ to be constant over the range $m_{c} \leq m_{H} \leq m_{b}$, the scaling relation in Eq.~(\ref{eq:scaling_width}) corresponds to the following scaling relation for the form factor $F_{1}(m_{H})$:
\begin{align}\label{eq:scalingF1}
F_{1}(m_{H}) \;\propto\; \frac{|\mathbf{k}|}{m_{H}\,v_{H}}\Bigl[1 + O\bigl(m_{H}^{-1}, v_{H}^{2}\bigr)\Bigr].
\end{align}
To extrapolate $F_{1}(m_{H})$ to the physical $b$-quark mass, we adopt an agnostic approach. This is made possible by the fact that our data are sufficiently precise and span over a large enough range of heavy meson masses (with the heaviest simulated $\eta_{H}$ being around $6.6\,\mathrm{GeV}$). Therefore any reasonable ansatz for the heavy mass dependence, that accurately describes the data in the simulated region, always yields results at the physical $b$-quark mass in the same ball park. By performing a sufficiently large number of fits we can then properly determine the systematic error associated with the mass extrapolation.

We employ the following forms for the extrapolation of the form factor:
\begin{itemize}
\item Type-A:  We exploit the fact that $\Delta m_{h\eta}$ is roughly constant when going from $m_{c}$ and $m_{b}$ and that parametrically $\Delta m_{h\eta}$ is of the order $O(m_{H} v_{H}^{2})$. We can therefore take $v_H \propto \sqrt{ |{\bf k}|/m_H}$ so that from Eq.~\eqref{eq:scalingF1} we get that $F_1(m_H) \propto 1/\sqrt{m_H}$,  up to $1/m_H$ corrections. Instead of the heavy quark mass one can use $m_{\eta_{H}}/2$, which is equivalent, up to higher order corrections in the velocity $v_{H}$. We can therefore use the form: 
\begin{align}\label{eq:scalingA}
F_1(m_H) \sqrt{m_{\eta_H}} = \alpha +{ \beta\over  m_{\eta_H}} + { \gamma\over  m_{\eta_H}^2}\,,
\end{align}
where $\alpha$, $\beta$ and $\gamma$ are the fit parameters, independent on the heavy quark mass. In Fig.~\ref{fig:sqrt_scaling} we show that the scaling~\eqref{eq:scalingA} is remarkably well verified by our lattice results. 
\begin{figure}
    \centering
    \includegraphics[width=0.8\linewidth]{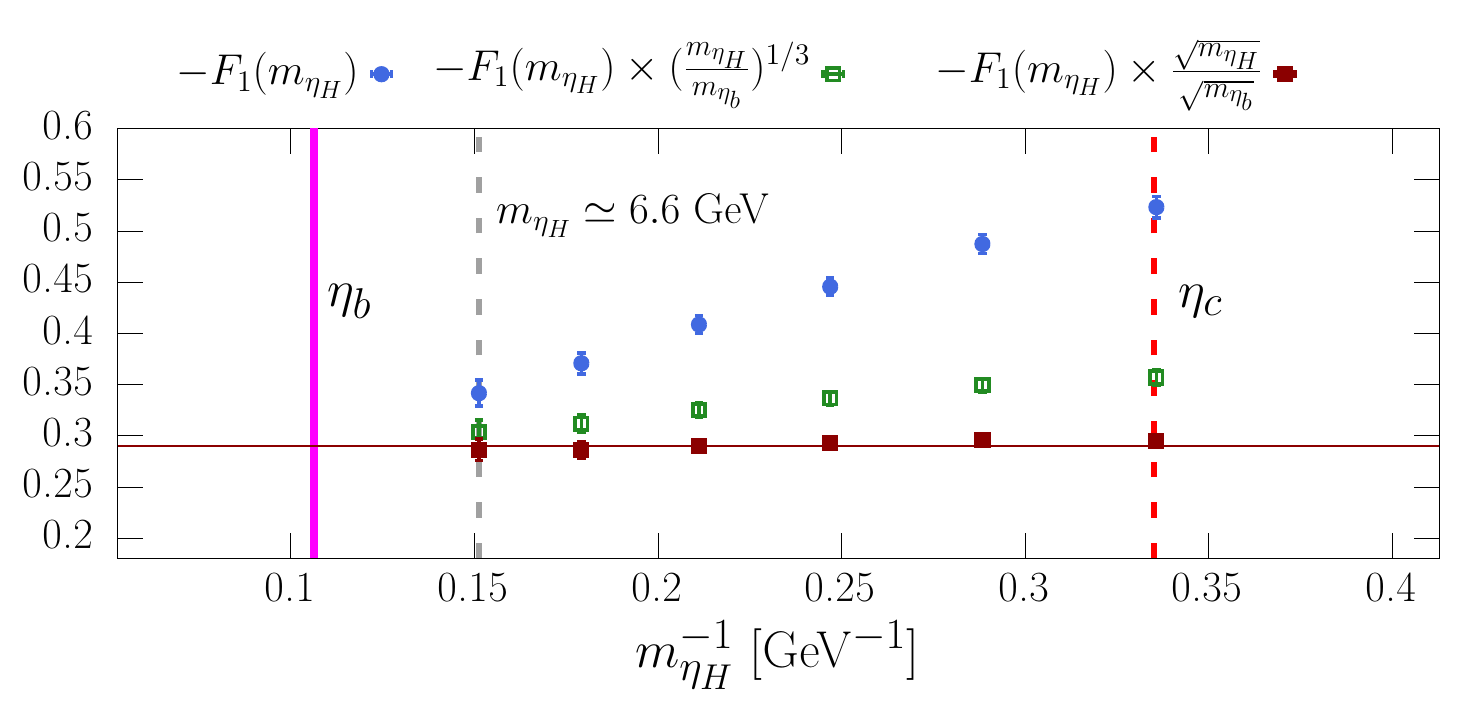}
    \caption{\small\sl In the figure we show our results for the form factor $F_{1}(m_{\eta_{H}})$ (blue circles) and for the combinations $F_{1}(m_{H}) \times (m_{\eta_{H}}/m_{\eta_{b}})^{1/3}$ (green empty squares) and $F_{1}(m_{H}) \times \sqrt{m_{\eta_{H}}}/\sqrt{m_{\eta_{b}}}$ (red filled squares). The horizontal line is added to guide the eye.}
    \label{fig:sqrt_scaling}
\end{figure}
From the fit of our data to Eq.~\eqref{eq:scalingA}, either by setting $ \widetilde \beta  =\widetilde{\gamma} = 0$, or by keeping them in the fit, we obtain 
\begin{equation}\label{eq:sqrt}
\text{Type-A}:\qquad F_1(m_b)\equiv F_1^b = -0.285(10)\,.
\end{equation}

\item Type-B: If the static limit is dominated by the Coulombic potential, as discussed above, one has  $v_{H} \propto \alpha_{s}\bigl(\frac{1}{2}m_{\eta_H}\,v_{H}\bigr)$~\cite{Brambilla:2012be,Segovia:2018qzb}, and by combining Eqs.~\eqref{eq:scaling0} with~\eqref{eq:scalingF1} we arrive at the fit form
\begin{align}\label{eq:scalingB}
F_1(m_H) = \alpha' +{ \beta'\over  m_{\eta_H}} + { \gamma'\over  m_{\eta_H}^2}\,,
\end{align}
where again, $\alpha'$, $\beta'$ and $\gamma'$ are the heavy quark mass independent fit parameters. 
To us these are merely fit parameters, while in pNRQCD the authors compute the $\alpha'$ term by subtracting renormalon and  using perturbative QCD to higher orders~\cite{Segovia:2018qzb}.  It is worth mentioning that from the linear fit to our data in $1/m_{\eta_H}$, we obtain $\alpha'_\mathrm{lin} = -0.20(2)$  [$\gamma'=0$ in  Eq.~\eqref{eq:scalingB}], while from the quadratic one we get $\alpha'_\mathrm{quad} = -0.17(6)$, where in both cases we removed our lightest heavy quark from the fit.
Again, by using the fit form~\eqref{eq:scalingB}, either by setting $  \beta'  =\gamma' = 0$, or by keeping them as free (fit parameters), we find 
\begin{equation} 
\text{Type-B}:\qquad  F_1^b = -0.298(20)\,.
\end{equation}

\item Type-C:  If instead of the Coulombic potential we assume a purely string potential $V_{H}(r) = \sigma\,r$, we get 
that $v_{H} \propto m_{\eta_{H}}^{-2/3}$, which then together with Eq.~\eqref{eq:scalingF1} lead to our third heavy quark mass dependence, 
\begin{align}\label{eq:scalingC}
F_1(m_H) \, m_{\eta_H}^{1/3} = \overline \alpha +{  \overline \beta\over  m_{\eta_H}} + {  \overline \gamma\over  m_{\eta_H}^2}\,.
\end{align}
After proceeding like in the previous two cases, we get
\begin{equation} 
\text{Type-C}:\qquad  F_1^b = -0.293(15)\,.
\end{equation}
\end{itemize}
In Fig.~\ref{fig:sqrt_scaling} we show the data and illustrate the quality of the scalings~\eqref{eq:scalingA} and~\eqref{eq:scalingB}, to emphasize the amount of corrections $\propto 1/m_{\eta_H}$ and higher.

\begin{figure}
    \centering
    \includegraphics[width=0.8\linewidth]{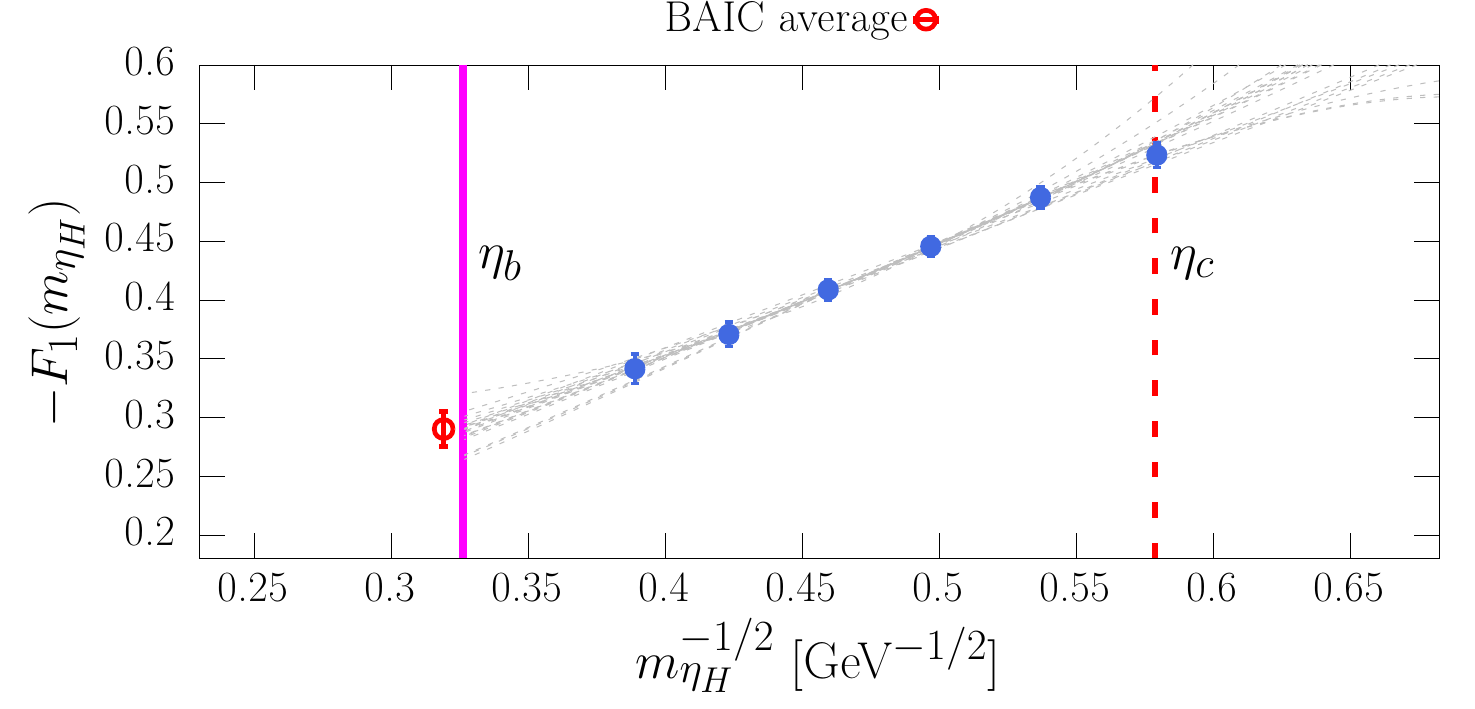}
    \caption{\small\sl Extrapolation of the form factor $F_{1}(m_{H})$ to $F_{1}(m_{b})$. Each of the black dashed curves corresponds to the best fit obtained in all of the fits we performed and discussed in the text. The red circle corresponds to the BAIC average of all fits, our final result. Since our results very well verify the scaling~\eqref{eq:scalingA}, we plot them as a function of $1/\sqrt{m_{\eta_H}}$.}
    \label{fig:extr_b}
\end{figure}

We fit our data using Eqs.~(\ref{eq:scalingA},\ref{eq:scalingB},\ref{eq:scalingC}) and for each type of fit we perform four different extrapolations: we first set the power corrections to zero and then include either the linear ($\propto 1/m_{\eta_H}$) or the quadratic term ($\propto 1/m_{\eta_H}^2$), and finally we include both of them. We also perform the fits by imposing cuts on the data which means that we repeated each of the above fits by excluding the the data point corresponding to the lightest simulated $m_{H}=m_{c}$, and even by dropping the results corresponding to the lightest two of our heavy quarks.  All fits have been performed by minimizing $\chi^{2}$ that fully takes into account the cross correlations among the data. 
All these fit results are then combined using the BAIC, which has been already discussed in Sec.~\ref{sec:num_results_hc}. In Fig.~\ref{fig:extr_b} we show the results of these extrapolations. The dashed lines correspond to the (central value of the) best fit curves, obtained in all the fits we performed (excluding  those leading to $\chi^{2}/\mathrm{dof} > 2$). In Fig.~\ref{fig:extr_b} we also show the final result obtained using the BAIC, namely,
\begin{align}
\boxed{~F_{1}^{b} = -0.290(15)\,,~}
\end{align}
which has an uncertainty of about $5\%$, three times larger than in the case of charmonia~\eqref{eq:F1c}. Note that this is the first lattice QCD result of this quantity.

\section{Comparison with existing theoretical and experimental results}
\label{sec:comparison}
In this Section we provide the decay widths $\Gamma(h_{c(b)} \to \eta_{c(b)} \gamma)$ and compare our results with existing experimental data and other lattice and non-lattice predictions.

Using the expression of Eq.~\eqref{eq:decay_rate_charm} and our result  for $F_{1}^{c} = -0.522(10)$, we obtain
\begin{align}
\label{eq:final_decay_rate_charmonium}
\Gamma(h_{c}\to\eta_{c}\gamma ) = (0.604 \pm 0.024)~\mathrm{ MeV}\,, 
\end{align}
where we used $\alpha_\mathrm{em}^{-1}(m_c) = 133.5(5)$~\cite{PDG2024}. Our result agrees with the experimental measurement by the BES-III Collaboration~\cite{BESIII:2016hfo,BESIII:2022tfo}: 
\begin{align}
\Gamma(h_{c}\to\eta_{c}\gamma)^\mathrm{exp} = (0.47 \pm 0.17)~\mathrm{ MeV}\,. 
\end{align}
although the current experimental uncertainty is much larger compared to ours. 
Note, however, that the experimental error is dominated by the total width, $\Gamma(h_{c})^\mathrm{exp} =( 0.78\pm 0.28)~\mathrm{ MeV}$~\cite{PDG2024}, which is about $35\%$, while the uncertainty on the branching fraction is relatively small ($7\%$). 
We can therefore combine our lattice QCD result~\eqref{eq:final_decay_rate_charmonium} with ${\mathcal{B}}(h_{c}\to \eta_{c} \gamma)^\mathrm{exp} = (60\pm 4)\%$\cite{PDG2024} and estimate $\Gamma(h_{c})$. We obtain 
\begin{align}
\Gamma(h_{c})^\mathrm{latt+exp} = \frac{\Gamma(h_{c}\to\eta_{c}\gamma)^\mathrm{latt} }{\mathcal{B}(h_{c}\to\eta_{c}\gamma)^\mathrm{exp} } = (1.007 \pm 0.078)~\mathrm{ MeV}\,.
\end{align}
This result has more than $4$ times smaller error that the experimentally measured total width~\cite{BESIII:2022tfo}.

\begin{figure}
    \centering
    \includegraphics[width=0.495\linewidth]{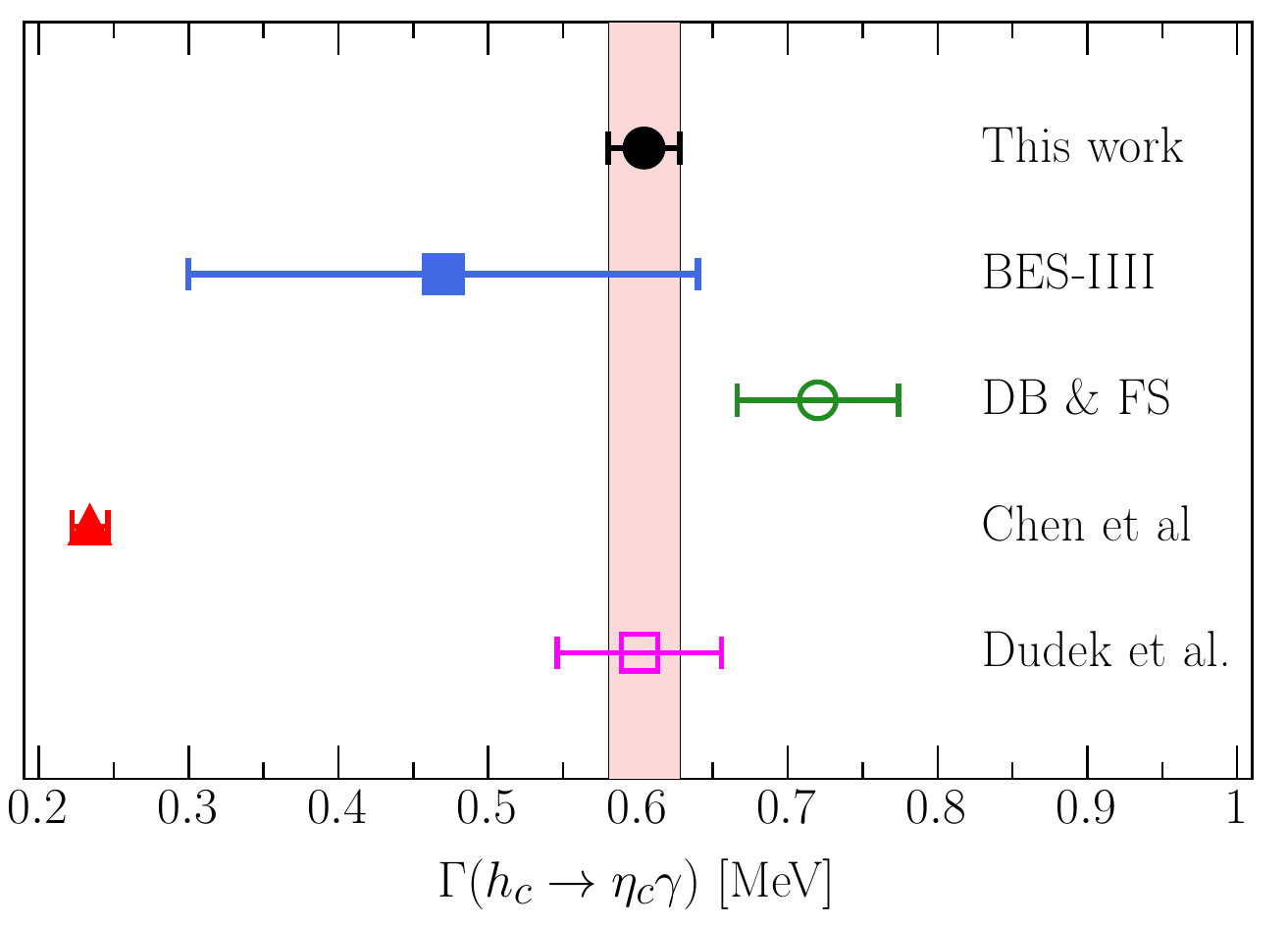}
    \includegraphics[width=0.495\linewidth]{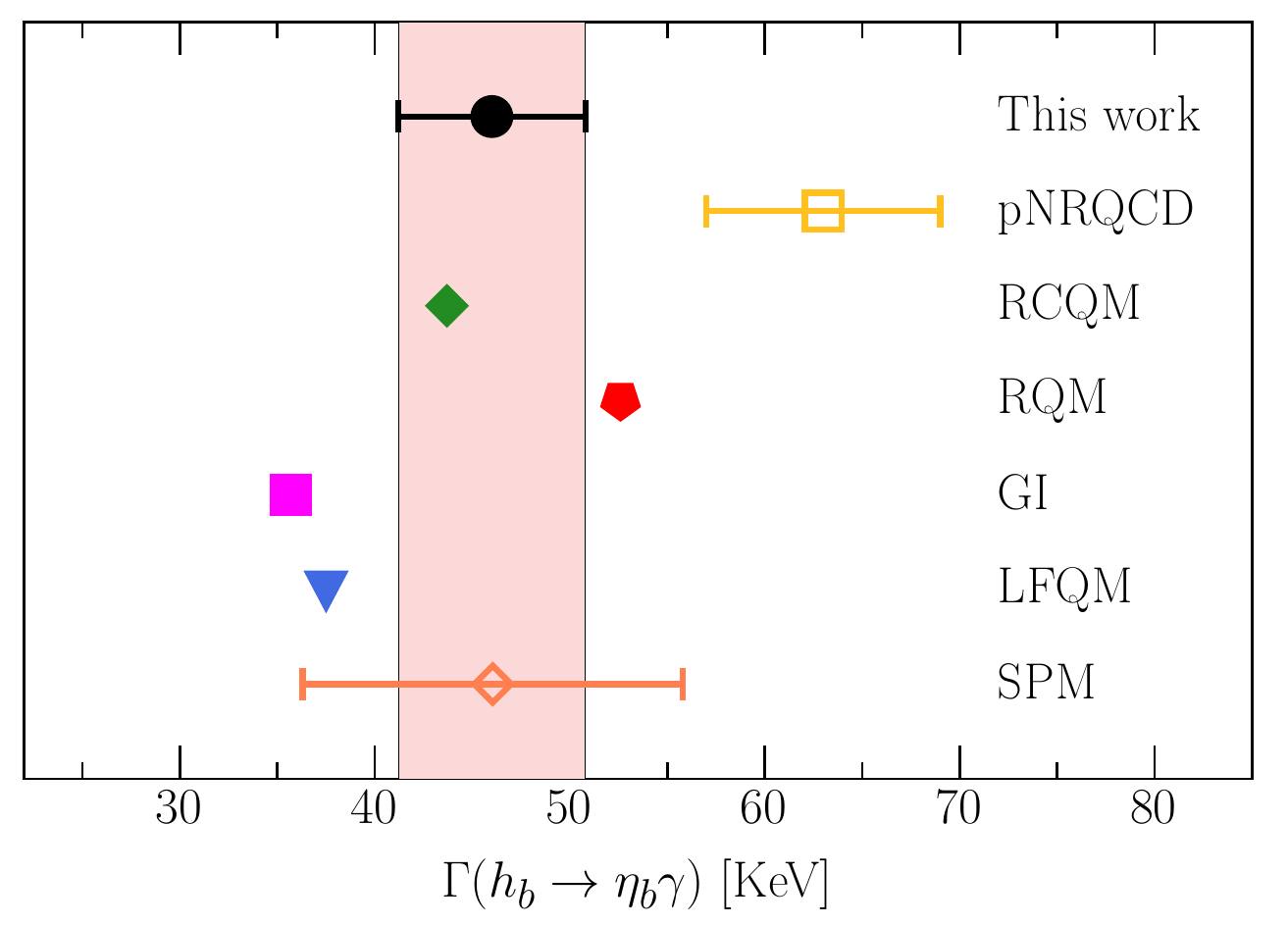}
   \caption{\small\sl Comparison of our estimate of the radiative decay widths 
$\Gamma\bigl(h_{c,b}\to\eta_{c,b}\gamma\bigr)$ (filled black circle) with results available in the literature. In the case of charm we compare with the BES-III measurement~\cite{BESIII:2016hfo,BESIII:2022tfo} (filled blue square), and the previous lattice results: Ref.~\cite{Becirevic:2012dc} (empty green circle),~\cite{Chen:2011kpa} (filled red triangle), and~\cite{Dudek:2006ej} (empty magenta square). In the right panel we compare our result for the bottomium with other approaches (labels with corresponding references can be found in the text, after Eq.~\eqref{eq:width-hb}).}
\label{fig:comp_decay_charm}
\end{figure}

Our $F_{1}^{c} = -0.522(10)$ is a substantial improvement over $F_{1}^{c} = -0.57(2)(1)$ of Ref.~\cite{Becirevic:2012dc} because it is obtained with $N_f=2+1+1$ dynamical quark flavors and (more importantly) with the physical sea quark masses. Their rate,  
$\Gamma(h_{c}\to\eta_{c}\gamma ) = 0.72(5)(2)$~MeV, is about $2\sigma$ larger than our result~(\ref{eq:final_decay_rate_charmonium}).~\footnote{Note that we corrected the expression for the matrix element in Ref.~\cite{Becirevic:2012dc}.} 
To the best of our knowledge there are only two other lattice QCD results. On the basis of a quenched lattice calculation, the rate $\Gamma(h_{c}\to\eta_{c}\gamma ) =  0.601(55)~{\rm MeV}$ has been reported in Ref.~\cite{Dudek:2006ej}, in agreement with our result~(\ref{eq:final_decay_rate_charmonium}). In Ref.~\cite{Chen:2011kpa}, instead, by using $N_{f}=2$  twisted-mass fermions at a single lattice spacing ($a\simeq 0.07~\mathrm{ fm}$) the authors quote $\Gamma(h_{c}\to\eta_{c}\gamma ) =  0.234(12)~\mathrm{ MeV}$, much smaller than our result. In Fig.~\ref{fig:comp_decay_charm} we show a comparison of different lattice results with existing model and experimental determinations of the radiative decay rate.

We now discuss $h_{b}\to\eta_{b}\gamma$. Like in the case of charmonia, we insert our result $F_{1}^{b
} = -0.290(15)$, in the expression of Eq.~(\ref{eq:decay_rate_heavy}) and obtain
\begin{align}
\label{eq:final_decay_rate_bottomonium}
\Gamma(h_{b}\to\eta_{b}\gamma) = (46.0 \pm 4.8)~\mathrm{keV}\,, 
\end{align}
where we used $\alpha_\mathrm{em}^{-1}(m_b) = 132.2(3)$~\cite{PDG2024}. Our result for the decay rate thus has an uncertainty of about $10\%$. In this case only the branching fraction has been measured $\mathcal{B}(h_{b}\to\eta_{b}\gamma)^\mathrm{exp} = 52^{+6}_{-5}\%$~\cite{Belle:2012fkf,Belle:2015hnh,PDG2024}. We can combine that result with our decay width estimate~\eqref{eq:final_decay_rate_bottomonium} and predict the total width of $h_{b}$. We get:
\begin{align}\label{eq:width-hb}
\Gamma(h_{b})^\mathrm{latt+exp} = \frac{\Gamma(h_{b}\to\eta_{b}\gamma)^\mathrm{latt} }{\mathcal{B}(h_{b}\to\eta_{b}\gamma)^\mathrm{exp} } = (88 \pm 13)~\mathrm{ keV}\,.
\end{align}
Since no other lattice QCD result relevant to $h_{b}\to\eta_{b}\gamma$ is available, we compare our estimate for the decay width~\eqref{eq:final_decay_rate_bottomonium} with the value obtained 
by using potential NRQCD~\cite{Segovia:2018qzb} and labeled as pNRQCD in Fig.~\ref{fig:comp_decay_charm}. We
also compare our result with various (relativistic) quark models which are labeled in Fig.~\ref{fig:comp_decay_charm} as RCQM~\cite{Segovia:2016xqb}, RQM~\cite{Ebert:2002pp}, GI~\cite{Godfrey:2015dia}, LFQM~\cite{Shi:2016cef}, SPM~\cite{Li:2009nr}. 
The overall agreement is good; the largest differences are with Ref.~\cite{Godfrey:2015dia}, which is $2\sigma$ below our result, and with the determination obtained using pNRQCD at weak coupling, which is a little over $2\sigma$ larger than our result. 

\section{Conclusions}
\label{sec:conclusions}
In this paper we provide the full QCD computation of the form factors relevant to the radiative decays $h_{c,b}\to\eta_{c,b}\gamma$. 

The result for the charmonium case is $F_1^c= - 0.522(10)$ is a very significant improvement over the previous lattice determinations because it is obtained with the physical sea quark masses with $N_f=2+1+1$ and with five fine lattices, allowing a smooth extrapolation to the continuum limit. For the reasons explained in the text, we dropped the disconnected diagrams in this calculation. Their impact can and should be studied separately. We plan to do so using the Hansen--Lupo--Tantalo method of Ref.~\cite{Hansen:2019idp} to overcome the difficulties of isolating the charmonium and bottomonium contributions in the quark-disconnected terms. Our final result for the corresponding decay width is $\Gamma(h_{c}\to\eta_{c}\gamma ) = (0.604 \pm 0.024)~\mathrm{ MeV}$.

We produced the first lattice QCD estimate of the form factor relevant to $h_{b}\to\eta_{b}\gamma$. Our  result $F_{1}^{b} = -0.290(15)$ is obtained through a heavy quark extrapolation of the results obtained by working with the charm quark and a series of quarks heavier than charm, up to three times the charm quark mass. After a controlled extrapolation to the continuum limit, we made extrapolation from the fictitious $F_1^H$ and $m_{\eta_H}$ to the desired (physical) $F_1^b$ and $\eta_b$. To do so we considered several options and performed many different fits. It should be emphasized that our results exhibit a very pronounced scaling of $F_1^H \sqrt{m_{\eta_H}} =$ constant, an observation that might be useful for pNRQCD studies. 
Like in the case of charmonia, also in this case we dropped the disconnected diagrams assuming their contribution to be negligibly small. From our estimate of the form factor $F_1^b$, we obtain $\Gamma(h_{b}\to\eta_{b}\gamma) = (46.0 \pm 4.8)~\mathrm{keV}$, which is also the first lattice QCD based determination of this quantity.

\section{Acknowledgments}
\label{sec:akno}
We thank the ETMC for the most enjoyable collaboration, N.~Brambilla and A.~Vairo for helpful communication regarding the extrapolation of the form factor to the $b$-quark mass, and S. Simula for useful discussions. V.L., F.S., G.G., R.F., and N.T. are supported by the Italian Ministry
of University and Research (MUR) and the European
Union (EU) – Next Generation EU, Mission 4, Component 1, PRIN 2022, CUP F53D23001480006. 
F.S. is supported by ICSC – Centro Nazionale di Ricerca in High Performance Computing, Big Data and Quantum Computing, funded by European Union - Next Generation EU and by Italian  Ministry of University and Research (MUR) projects FIS\_00001556 and PRIN\_2022N4W8WR. We acknowledge support from the LQCD123, ENP, and SPIF Scientific Initiatives of
the Italian Nuclear Physics Institute (INFN). 
This project has received support from the European Union’s Horizon 2020 research and innovation programme under the
Marie Sklodowska-Curie grant agreement N◦ 860881-HIDDeN and N◦ 101086085-ASYMMETRY and the IN2P3 (CNRS) Master Project HighPTflavor.

The open-source packages tmLQCD~\cite{Jansen:2009xp,Abdel-Rehim:2013wba,Deuzeman:2013xaa,Kostrzewa:2022hsv}, LEMON~\cite{Deuzeman:2011wz}, DD-$\alpha$AMG~\cite{Frommer:2013fsa,Alexandrou:2016izb,Bacchio:2017pcp,Alexandrou:2018wiv}, QPhiX~\cite{joo2016optimizing,Schrock:2015gik} and QUDA~\cite{Clark:2009wm,Babich:2011np,Clark:2016rdz} have been used in the ensemble generation.

We gratefully acknowledge the ICSC - Centro Nazionale di Ricerca in High Performance Computing for providing computing time under the allocations RAC 1916318. We gratefully acknowledge CINECA for the provision of GPU time on Leonardo supercomputing facilities under the specific initiative INFN-LQCD123, and under project IscrB VITO-QCD and project IscrB SemBD. We gratefully acknowledge EuroHPC Joint Undertaking for awarding us access to MareNostrum5 through the project EHPC-EXT-2024E01-031. The authors gratefully acknowledge the Gauss Centre for Supercomputing e.V. (www.gauss-centre.eu) for funding this project by providing computing time on the GCS Supercomputers SuperMUC-NG at Leibniz Supercomputing Centre. The authors acknowledge the Texas Advanced Computing Center (TACC) at The University of Texas at Austin for providing HPC resources (Project ID PHY21001). We gratefully acknowledge PRACE for awarding access to HAWK at HLRS within the project with Id Acid 4886. We acknowledge the Swiss National Supercomputing Centre (CSCS) and the EuroHPC Joint Undertaking for awarding this project access to the LUMI supercomputer, owned by the EuroHPC Joint Undertaking, hosted by CSC (Finland) and the LUMI consortium through the Chronos programme under project IDs CH17-CSCS-CYP. We acknowledge EuroHPC Joint Undertaking for awarding the project ID EHPC-EXT-2023E02-052 access to MareNostrum5 hosted by at the Barcelona Supercomputing Center, Spain.

\bibliography{biblio}

\end{document}